\documentclass[%
 reprint,
 nofootinbib,
 amsmath,amssymb,
 pra,
]{revtex4-1}

\usepackage{graphicx}
\usepackage{dcolumn}
\usepackage{braket}
\usepackage{bm}

\newcommand{\pd}{\partial}

\begin{document}

\title{Multi-level perspective on high-order harmonic generation in solids}

\author{Mengxi Wu}
\author{Kenneth J. Schafer}
\author{Mette B. Gaarde}
\email{gaarde@phys.lsu.edu}

\affiliation{Department of Physics and Astronomy, Louisiana State University, Baton Rouge, LA 70803-4001, USA}
\date{\today}

\begin{abstract}

We investigate high-order harmonic generation in a solid, modeled as a multi-level system dressed by a strong infrared laser field. We show that the cutoff energies and the relative strengths of the multiple plateaus that emerge in the harmonic spectrum can be understood both qualitatively and quantitatively by considering a combination of adiabatic and diabatic processes driven by the strong field.
Such a model was recently used to interpret the multiple plateaus exhibited in harmonic spectra generated by solid argon and krypton [Ndabashimiye {\it et al.}, Nature 534, 520 (2016)]. 
We also show that when the multi-level system originates from the Bloch state at the $\Gamma$ point of the band structure, the laser-dressed states are equivalent to the Houston states [Krieger {\it el al.} Phys. Rev. B 33, 5494 (1986)] and will therefore map out the band structure away from the $\Gamma$ point as the laser field increases. This leads to a semi-classical three-step picture in momentum space that describes the high-order harmonic generation process in a solid. 

\end{abstract}

\maketitle

\section{Introduction}

Beginning with its first observation in rare gases almost 30 years ago \cite{Ferray1988}, high-order harmonic generation (HHG) has become the foundation for attosecond science through a series of advances at the fundamental as well as the applied level \cite{Farkas1992, Harris1993, Schafer1993, Corkum1993, Brabec2000, Sansone2006, Krausz2009, Wright2010}. 
Since the turn of the  millennium, attosecond pulses produced via gas-phase HHG have become a successful tool for the study of ultrafast dynamics in atoms, molecules, and biological systems \cite{Hentschel2001, Smirnova2009, Worner2010, Morishita2008}. The recent observation of HHG from bulk solids, displaying a plateau of harmonics that ends in a high frequency cutoff~\cite{Ghimire2010, Ghimire2011, Schubert2014, Hohenleutner2015, Mahmood2015,Vampa2015a, NdabashimiyeG.2016a}, has generated considerable interest given its potential as both a compact, next generation  ultrafast light source in the extreme ultraviolet (XUV), as well as the promise of applying HHG spectroscopic techniques to correlated electron dynamics in bulk materials \cite{Hohenleutner2015}.

The mechanism for HHG in solids has been intensely debated \cite{Korbman2013, Kemper2013, Vampa2014, Higuchi2014a, Hawkins2015, Wu2015, Vampa2015b, Vampa2015, McDonald2015, Hawkins2015, Guan2015}, and a conceptual picture of the generation process, augmented by new experimental results~\cite{NdabashimiyeG.2016a,Vampa2015c,Hohenleutner2015,Vampa2015a,Schubert2014}, is only slowly beginning to emerge. 
In momentum space, the discussion has centered on the relative contributions of inter-band and intra-band processes to the driven non-linear current that gives rise to the harmonic radiation \cite{Ghimire2010, Vampa2014, Schubert2014, Higuchi2014a, Wu2015, Hawkins2015}. A consensus is forming that while intra-band contributions are important for harmonics with energies below the band gap \cite{Schubert2014}, inter-band processes generally dominate the higher harmonics that span the plateau region \cite{Vampa2015b, Wu2015}. In real space, Vampa and collaborators~\cite{Vampa2014} have proposed a three-step semi-classical model in which a localized electron and hole pair are accelerated away from and recollide with each other after traversing many lattice cells, in analogy with the gas-phase three-step model of HHG~\cite{Schafer1993, Corkum1993}. 

While most of the works discussed above have described the strong-field dynamics using only a valence band and the lowest-lying conduction band, recent experimental and theoretical findings indicate another layer of complexity in the HHG generation process: Ndabashimiye {\it et al.}~\cite{NdabashimiyeG.2016a} directly compared the harmonics from argon and krypton in their solid phase and their gas phase and found that the solid HHG spectra exhibited multiple plateaus with the highest cutoff energies far exceeding those found in the gas phase for the same laser intensity. The multiple plateaus and their relative strengths and cutoff energies were reproduced using a multi-level model that takes into account the coupling between the valence band and several conduction bands at the high symmetry ($\Gamma$) point in solid argon \cite{NdabashimiyeG.2016a}. 

In this paper we explore in detail the multi-level model used in \cite{NdabashimiyeG.2016a} to calculate the HHG spectrum from a solid interacting with a strong, mid-infrared laser field. We show that the appearance of multiple plateaus in the harmonic spectrum from such a system can be understood both qualitatively and quantitatively, and we derive predictions for the cutoff energies and the relative strengths of the different plateaus. We begin by showing that a multi-level system is a natural basis in which to consider harmonic generation in a solid, because when  the time-dependent Schr\"odinger equation (TDSE) in a periodic system is solved in the velocity gauge the different values of the crystal momentum $k_0$  remain uncoupled even in the presence of a laser field~\cite{Krieger1986, Korbman2013}. The band structure of the solid can thus be represented as a collection of independent multi-level systems, and the dynamics in the solid is given by the sum of the dynamics of each $k_0$-value represented in the initial wave function. Here we concentrate on the dynamics that result from the simplest possible initial condition, in which a single Bloch state with $k_0=0$ ($\Gamma$ point) is considered so that the initial electron wave function is maximally delocalized. In \cite{Wu2015} we showed that this very simple condition reproduces the characteristics of inter- and intra-band contributions to HHG and their relative importance to low and high order harmonics. 
We interpret the electron dynamics in the multi-level system in terms of the time-dependent eigenstates of the laser-dressed system, the adiabatic states, and we show how  HHG  proceeds through a combination of adiabatic and diabatic processes involving the dressed states, similarly to what has been discussed for instance in \cite{Higuchi2014a}. 
We use the evolution of the adiabatic states to derive predictions for the cutoff energies and relative strengths of each plateau supported by the multi-level system. 

After discussing HHG in multi-level systems, we show that when the multi-level system is constructed from the energy levels and their transition matrix elements for a periodic potential at $k_0=0$, the adiabatic states will recover the full $k_0 \ne 0$ band structure of that solid, if we allow the crystal momentum to increase with the vector potential of the field, similar to what is done in the Houston state treatment of reference~\cite{Krieger1986}. This correspondence between the adiabatic states and the band structure leads us to a semi-classical, three-step picture for harmonic generation in a solid, in momentum space. In this picture, the delocalized electron first tunnels from the valence band (VB) to the conduction band (CB) at the zero of the vector potential and is then accelerated on the conduction band as the vector potential increases and decreases through a half optical cycle. The coherence between the VB and CB populations leads to emission of XUV radiation, with photon energies corresponding to the instantaneous energy difference between the VB and the CB. This means that every energy below the cutoff energy is emitted twice in each half-cycle.  
This picture also allows us to predict that for a 1D multi-band system, the maximum cutoff energy of each plateau will be limited by the maximum band gap between the valence band and the highest-lying conduction band involved in producing that plateau. This extends the cutoff limitation proposed in \cite{Vampa2015} and \cite{Higuchi2014a} for two-band systems to a multi-band system exhibiting multiple plateaus in the harmonic spectrum.

The paper is organized as follows: In Section \ref{sec:TDSEsolid} we introduce the theoretical framework that the remainder of the paper is based on, namely the solution of the single-active-electron TDSE in a one-dimensional (1D) solid interacting with a strong field, and we discuss our initial condition of a delocalized electron wavefunction. 
In Section \ref{sec:two_level_HHG} we use the simplest multi-level system -- the two-level system -- to review the formalism for how to think about HHG in a multi-level system both at the qualitative level, using a three-step picture, and at the quantitative level in terms of predictions for the strength and extent (cutoff energy) of the plateau. In Section \ref{sec:HHG_multi-level_system} we study HHG in a multi-level system consisting of four or more levels, which, as discussed above, leads to the appearance of multiple plateaus in the harmonic spectrum. We extend the two-level formulas and derive expressions for the cutoff energies and relative strengths of the different plateaus. Finally, in Section \ref{sec:Semiclassical_picture} we show the connection between the adiabatic states of the dressed multi-level system and the band structure of the model solid and the resulting semi-classical picture of HHG in solids. Section \ref{sec:Summary} presents a brief summary of our results. We will use  atomic units through out this paper.

\section{Solving the TDSE for a 1D solid}
\label{sec:TDSEsolid}

Strong-field processes in solids are often modeled by solving the TDSE in the velocity gauge using the dipole approximation \cite{Korbman2013, Higuchi2014a, Wu2015, Hawkins2015}:
\begin{equation}
i\frac{\pd}{\pd t}\ket{\psi(t)}=\left(\frac{\hat{ p}^2}{2} + V(\hat{x})+\vec{A}(t) \cdot \hat{ p}\right)\ket{\psi(t)},
\label{Eq:TDSE_velocity}
\end{equation}
where $V(\hat{x})$ is a periodic, one-electron pseudo-potential that can be calculated approximately, for instance from density functional theory. $\vec{A}(t)$ is the vector potential of the laser pulse in the dipole approximation, where we assume that the laser wavelength (often one micron or greater) is much larger than the lattice constant, meaning we can take $\vec{A}$ to be coordinate independent. Throughout the paper, we will consider a 1D model, so that the $\hat{x}$ and $\vec{A}$ reduce to 1D quantities. This models a thin crystal with one of the transverse crystal directions (for example the $\Gamma-X$) aligned with the laser polarization. 

When solving Eq.~\eqref{Eq:TDSE_velocity} in the independent particle model it is convenient to expand the time-dependent wave function in the eigenstates of the field free Hamiltonian, {\it i.e.}, the Bloch-state basis
\begin{align}
\ket{\psi_{k_0}(t)} &= \sum_n C_{nk_0}(t) \ket{\phi_{nk_0}},
\label{eqn:expand_wave_function_in_Bloch_states}
\end{align}
where $\ket{\phi_{nk_0}}$ is the Bloch state for a specific crystal momentum $k_0$ in the $n^{th}$ band. This expansion allows us to take advantage of the fact that within the dipole approximation each crystal momentum channel $k_0$ is independent \cite{Korbman2013}, even in the presence of the field. 
After  expanding in the Bloch state basis, the TDSE in each crystal momentum channel is
\begin{equation}
i\frac{\partial}{\partial t} C_{nk_0} = C_{nk_0} \varepsilon_n(k_0) + A(t)\sum_{n'} C_{n'k_0} p^{nn'}_{k_0},
\label{eqn:TDSE_in_Bloch_basis}
\end{equation}
where $\varepsilon_n(k_0)$ is the energy of the $n^{th}$ band at $k_0$, and $p^{nn'}_{k_0}$ is the transition matrix element between the Bloch states with the same $k_0$ and different band indices:
\begin{align}
p^{nn'}_{k_0}&=\braket{\phi_{nk_0}|\hat p|\phi_{n'k_0}}.
\end{align}
We note here that Eq.~\eqref{eqn:TDSE_in_Bloch_basis} for a single $k_0$ channel is simply the TDSE for a multi-level system interacting with a laser field, so that the solid dynamics can be constructed as an ensemble of independent multi-level systems.

To calculate the HHG signal we need the coherent part of the current, which is
\begin{align}
 j_{k_0}(t)  &= -\left[\braket{\psi_{k_0}(t)|\hat p|\psi_{k_0}(t)}+A(t)\right]\\[0.5em]
          &= - \left[\text{Re}\left[\sum_{n,n'} C^*_{nk_0}(t)C_{n'k_0}(t)p^{nn'}_{k_0}\right]+A(t)\right]\nonumber
\label{Eq:total_current_Bloch_basis}
\end{align}
in each $k_0$ channel. In principle, we need to consider all possible $k_0$ in the first Brillouin zone and calculate the total current as
\begin{equation}
    j(t) = \int\limits_{k_0\in\text{BZ}} j_{k_0}(t) dk_0. 
\end{equation}
In this paper, however, we will restrict our study to the single excitation channel, $k_0=0$, corresponding to a maximally (spatially) delocalized Bloch state at the high symmetry point $\Gamma$ of the band structure. In our 1D model, as it is in many direct band gap materials, the transition matrix element between the VB and CB is maximized at $k_0=0$. This is in particular true in solid argon which we will be using as an example below. This means that the electron at $k_0=0$ has the highest excitation probability to the higher conduction bands. One therefore expects the largest contribution to the highly nonlinear HHG yield to originate at $k_0=0$ as has indeed been verified in calculations incorporating the 3D band structure \cite{Vampa2015b}, at moderate intensities. 
Moreover, for the purposes of this work, using a single $k_0$ channel allows us a straight-forward extension beyond a two-band model which was shown to be necessary to reproduce the multiple harmonic plateaus observed in recent experiments \cite{Wu2015,NdabashimiyeG.2016a}. 
With this initial condition, the dynamics of the solid is equivalent to that of a multi-level system, where the levels are the Bloch states for the different bands at a specific $k_0$, and the transition matrix elements are those between the Bloch states of the same $k_0$ in different bands.

\section{High harmonic generation in a two-level system}
\label{sec:two_level_HHG}

In this section, we review HHG in the  simplest multi-level system -- the two-level system. The HHG process in a two-level system has been studied extensively before, mostly in the context of atomic states with a coupling specified in terms of the electric field, in the length gauge \cite{KraTnov1980,Averbukh1985,Plaja1992,Ivanov1993,Krainov1994,Gauthey1997a,Santana2000,FigueiradeMorissonFaria2002,Haljan2003,Faria2003,Ashhab2007,Pic??n2010,Mkrtchian2012,Garraway1992}. 
Our two-level system originates in a two-band solid and  consists of Bloch states in two different bands but the same $k_0$ that are coupled by the vector potential as shown in Eq.~\eqref{eqn:TDSE_in_Bloch_basis}. The Hamiltonian for our laser-driven two-level system can then be written as
\begin{equation}
H=\left(
\begin{array}{cc}
 \omega _1 & \mu A(t) \\
 \mu A(t)  & \omega_2 \\
\end{array}
\right),
\end{equation}
where the two levels correspond to the valence and conduction band at $k_0=0$. $\mu=p^{vc}_{k_0=0}$ is the transition matrix element between these two Bloch states,  $\omega_1$ and $\omega_2$ are the energy of the highest valence band and the lowest conduction band, respectively. Their energy difference is defined as $\omega_0 = \omega_2-\omega_1$, which is the band gap at $k_0=0$. 
In our numerical calculations we use a laser pulse with a central frequency $\omega$. The pulse is derived from the vector potential $A(t)$ that has a $\cos^4$ envelope:
\begin{equation}
A(t) = A_0 \cos^4\left(\frac{\omega t}{2n}\right) \cos(\omega t), ~~~-\frac{n\pi}{\omega}\leq t \leq \frac{n\pi}{\omega}.
\end{equation}
$A_0$ is the peak vector potential and $n$ is the total number of cycles used in the calculation. It's also useful to define the peak Rabi frequency $\Omega_0$ and the time-dependent Rabi frequency $\wp(t)$:
\begin{align}
\Omega_0 & = \mu A_0\\
\wp(t) &= \mu A(t).
\end{align}
Then the Hamiltonian for this two-level model can be written as
\begin{equation}
H=\left(
\begin{array}{cc}
 \omega _1 & \wp(t) \\
 \wp(t)  & \omega_2 \\
\end{array}
\right).
\label{eqn:two_level_Hamiltonian}
\end{equation}

\begin{figure}[tb!]
\centering
\includegraphics[width=0.5\textwidth]{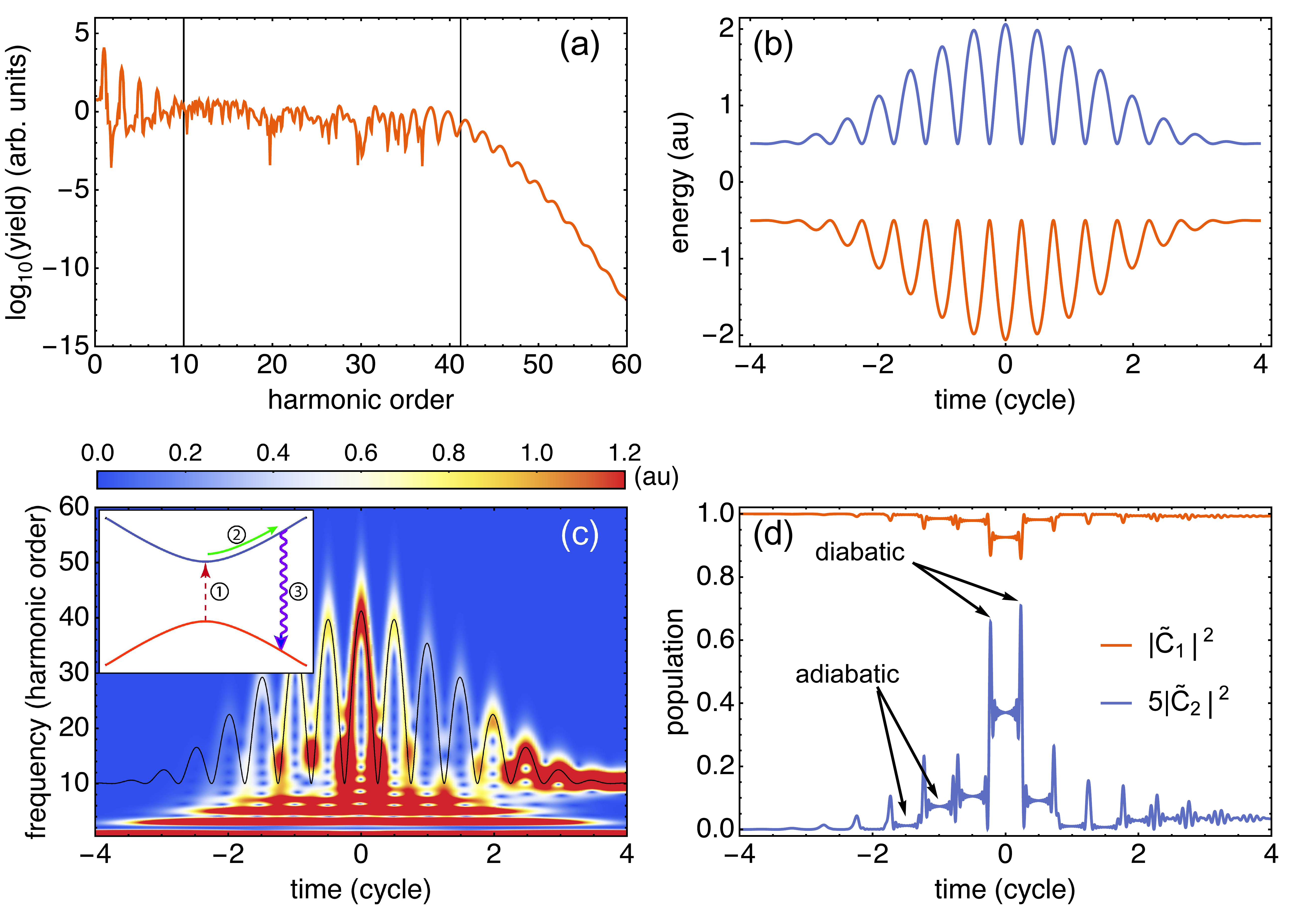}
\caption{(Color online) (a) Harmonic spectrum for a two-level system, with parameters $\omega_0=1$ au, $\Omega_0=2$ au and $\omega=0.1$ au. (b) Adiabatic state energies as a function of time. Avoided crossings are formed at the zeroes of the vector potential. (c) Wavelet transform of the time-dependent current, corresponding to the emission time of the harmonics. The thin black line indicates the energy difference between the adiabatic states in (b). The inset shows the three-step picture for the HHG process near an avoided crossing (see text). (d) Population on the two adiabatic eigenstates. }
\label{fig:two_level_HHG}
\end{figure}

A typical harmonic spectrum from a strongly driven two-level system is shown in Fig.~\ref{fig:two_level_HHG}(a), where the two-level energy separation is  $\omega_0 = 1$ au, the laser frequency is $\omega = 0.1$ au, the Rabi frequency is $\Omega_0 = 2$ au, and the total number of cycles in the pulse is $n=11$, which is about 3 cycles in FWHM for the envelope we are using.
The spectrum exhibits a perturbative regime (photon energies between $\omega$ and $9\omega$), a plateau regime (photon energies between $9\omega$ and $41\omega$) and a cutoff regime with a fast decline  (photon energies $> 41\omega$).

For the parameters used in Fig.~1 the HHG process in the two-level system can  best be described using the time-dependent adiabatic states of the system, which are the instantaneous eigenstates of the time-dependent Hamiltonian in Eq.~\eqref{eqn:two_level_Hamiltonian}  \cite{Krainov1994,Gauthey1997a, FigueiradeMorissonFaria2002}. For the full description for the adiabatic basis, see Appendix A.
Diagonalizing the time-dependent Hamiltonian in Eq.~\eqref{eqn:two_level_Hamiltonian} we get the time-dependent energies of the adiabatic states
\begin{align}
\mathcal{E}_{\pm}(t) = \pm\sqrt{\wp^2(t) + \left(\frac{\omega_0}{2}\right)^2},
\label{eqn:two_level_adiabatic_energy}
\end{align}
which are shown in Fig.~\ref{fig:two_level_HHG}(b). For our choice of parameters, $\Omega_0=2\omega_0$, the dressed state energies $\mathcal{E}_{\pm}(t)$ almost trace the magnitude of the vector potential $\pm |A(t)|$. Their separation is maximized at the peaks of the vector potential, and minimized at the zeroes of the vector potential where the two adiabatic states form avoided crossings.

To illustrate the usefulness of the adiabatic basis, we explore the time-frequency properties of the two-level harmonics in Fig.~\ref{fig:two_level_HHG}(c), which shows the wavelet transform of the time-dependent current, using a Gabor wavelet $g(t)=\frac{1}{\sqrt[4]{\pi }}e^{-\frac{t^2}{2}+10 i t}$ \cite{Mathematica}. The wavelet transform is similar to a time-limited, sliding-window Fourier transform and reveals the emission time of the harmonics. The thin curve overlaid on top is the energy difference of the two adiabatic states $\mathcal{E}_+(t) - \mathcal{E}_-(t)$. We see that the two-level harmonic emission process involves two symmetric ``paths", and the emission time of each path agrees very well with the the adiabatic state energies.%

The time evolution of the excited state population is also most easily understood in the adiabatic state basis, as  shown in Fig.~\ref{fig:two_level_HHG}(d). In this basis, the population evolution can be separated into {\it adiabatic} and {\it diabatic} processes, as indicated by arrows in Fig.~\ref{fig:two_level_HHG}(d). The adiabatic process manifests itself as the dressed level populations staying the same, while in the diabatic process the population changes abruptly. The diabatic processes take place at the zeroes of the vector potential when the two states form avoided crossings and the time-dependent transition matrix element between the adiabatic states is greatly peaked (see Eq.~\eqref{eqn:X_matrix_two-level} in Appendix A). Between the avoided crossings the system evolves almost adiabatically, so the populations on the two adiabatic states stay almost the same. This adiabatic evolution corresponds to rapid oscillations in the bare state populations (not shown), and the resulting non-linear current is responsible for the high harmonics generated between the zeroes of the vector potential as shown in Fig.~\ref{fig:two_level_HHG}(c).

From the above analysis of the time-frequency properties of the harmonics and the population evolution on the adiabatic states, a three-step picture emerges for HHG in the two-level system in the adiabatic basis, as shown in the inset of Fig.~\ref{fig:two_level_HHG}(c). The inset shows the enlargement of an avoided crossing in Fig.~\ref{fig:two_level_HHG}(b). In the first step, the population tunnels through the avoided crossings from the lower adiabatic state to the upper adiabatic state. In the second step, the population on the adiabatic states evolves adiabatically, gaining a phase proportional to their time-dependent energy separation. In the final step, the coherence between the adiabatic states generates high harmonics with energies corresponding to the instantaneous energy separation between the two adiabatic states. The lowest harmonic generated in this process is at the bare state two-level energy difference\footnote{Harmonics with lower energies are generated in a perturbative process, with strengths that decrease rapidly with order}, and the highest harmonic is generated at the largest energy difference of the two dressed states \cite{Krainov1994,Gauthey1997a}
\begin{align}
E_\text{low} &= \omega_0\\
E_\text{cutoff} &= 2\sqrt{\Omega_0 ^2 + \left(\frac{\omega_0}{2}\right)^2}.
\label{eqn:two_level_cutoff_energy}
\end{align}
The lowest and the highest harmonics predicted by these formulas are indicated by the vertical lines in Fig.~\ref{fig:two_level_HHG}(a), and can be seen to agree very well with the extent of the plateau region.

The three-step picture of HHG in a two level system works well when the diabatic interactions are short compared to the adiabatic evolution, so that the ``diabatic tunneling" and the ``adiabatic following" can be separated in time. The condition for the well-separated adiabatic and diabatic processes can be written as a ``crossing parameter" $R=\omega_0/2\Omega_0$ \cite{Gauthey1997a}. For high intensity and small two-level separation such that $R \ll 1$, the adiabatic and diabatic processes are well-separated and thus the three-step picture works very well, whereas for low intensity and large two-level separation such that $R \gg 1$, the adiabatic and diabatic processes are not well-separated and the three-step picture may not apply. 
For our parameters in Fig.~\ref{fig:two_level_HHG}, $R=0.25$ and the adiabatic and diabatic description works very well as expected.

\begin{figure}[h]
\centering
\includegraphics[width=0.48\textwidth]{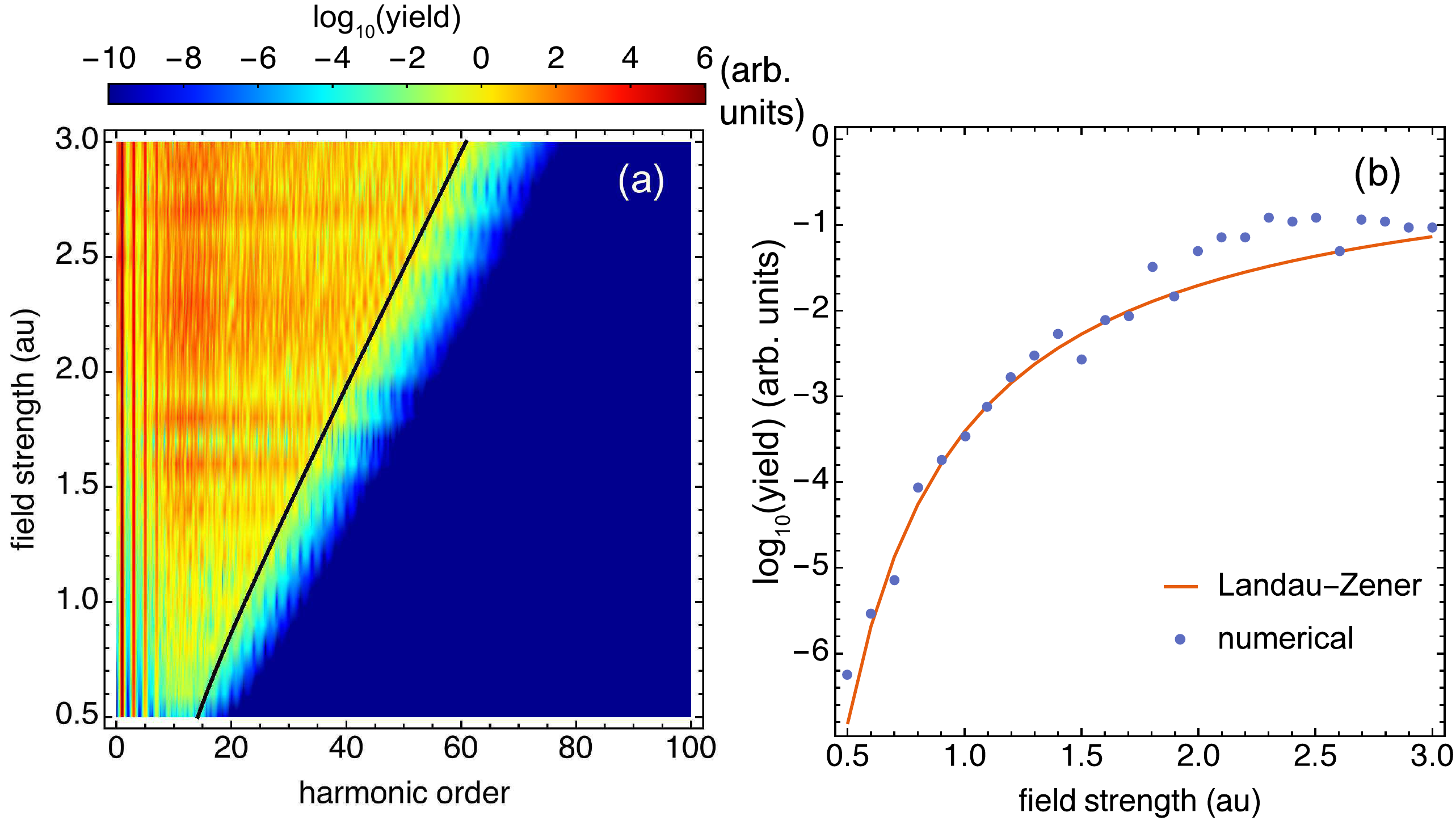}
\caption{(Color online) (a) Evolution of the harmonic spectrum as a function of the peak field strength. The solid line indicates the prediction for the cutoff from Eq.~\eqref{eqn:two_level_cutoff_energy}. As the Rabi frequency exceeds the two-level energy, the cutoff is approximately linear with the field strength. (b) The strength of the harmonic plateau (blue dots) agrees with the prediction of Landau-Zener tunneling (red solid line) very well.} 
\label{fig:two_level_HHG_intensity_scan}
\end{figure}

In the two-level model of harmonic generation, the cutoff is determined by the maximum energy difference between the adiabatic states, see Eq.~\eqref{eqn:two_level_cutoff_energy}. For intense fields, the cutoff is approximately linear with the field strength, {\it i.e.}, $E_\text{cutoff}  \approx 2 \Omega_0$. Fig.~\ref{fig:two_level_HHG_intensity_scan}(a) shows the field dependence of the harmonic spectrum, overlaid on top of which is the cutoff predicted from Eq.~\eqref{eqn:two_level_cutoff_energy}. The cutoff formula Eq.~\eqref{eqn:two_level_cutoff_energy} works very well. Since the harmonics are generated by the coherence between the adiabatic states, the harmonic strength is then determined by the population at the adiabatic states, which is directly related to the tunnelling rate between the adiabatic states at the avoided crossings. This tunneling rate can be calculated using the Landau-Zener tunneling formula near an avoided crossing \cite{Gauthey1997a}

\begin{equation}
P=\alpha\exp \left(-\frac{\pi  \omega _0^2}{4 \omega  \Omega _0}\right),
\label{eqn:Landau-Zener_tunneling_rate}
\end{equation}
where $\alpha$ is a scaling factor. The dots in Fig.~\ref{fig:two_level_HHG_intensity_scan}(b) show the yield of the harmonics in the plateau (measured by the strength of the cutoff harmonic) for different electric field values (defined as $E(t) = -\dot{A}(t)$). The solid line shows the prediction from Eq.~\eqref{eqn:Landau-Zener_tunneling_rate} with $\alpha=10$. We can see that Eq.~\eqref{eqn:Landau-Zener_tunneling_rate} works well in predicting the intensity of the plateau.

\section{HHG in a multi-level system}
\label{sec:HHG_multi-level_system}

In this section we expand from a two-level system to a multi-level system. We begin with a four-level system, again formed by the Bloch states of a 1D model solid at $k_0=0$. Our model solid has a Mathieu-type periodic potential $V(x) = -V_0\left[1+\cos(2\pi x/a_0)\right]$, with $V_0=0.37$ au and lattice constant $a_0=8$ au.
This potential conveniently has a band structure that can be expressed in terms of Matheiu characteristic values \cite{Slater1952}. The harmonic spectrum from this solid system can be calculated by solving the TDSE in the Bloch state basis
and the resulting harmonic spectrum exhibits multiple plateaus, as described in \cite{Wu2015}. The energies and coupling matrix elements for the lowest five Bloch states at $k_0=0$ can be calculated by solving the time-independent Schr\"odinger equation and are listed in Tab.~\ref{tab:five_level_system_parameters}. For our four-level system, we discard the lowest-energy Bloch state and use the second-lowest state as the valence band. Tab.~\ref{tab:five_level_system_parameters} also shows that the coupling between the four levels predominantly takes place in a stepwise manner, since the coupling matrix elements between neighboring levels are much larger than across multiple levels. 

\begin{table}[t]
\centering
\begin{tabular}{c|ccccc|c}
\hline
\multicolumn{6}{c|}{transition matrix element (au) (Eq.~(4))} & energy (au) \\ \hline
level     & 0    & 1    & 2    & 3    & 4    &       \\\hline
0        & 0    & 0.41 & 0    & 0.03 & 0    & -0.526\\
1        & 0.41 & 0    & 0.70 & 0    & 0.14 & -0.098\\
2        & 0    & 0.70 & 0    & 0.18 & 0    &  0.056\\
3        & 0.03 & 0    & 0.18 & 0    & 1.55 &  0.878\\
4        & 0    & 0.14 & 0    & 1.55 & 0 &  0.880\\\hline
\end{tabular}
\caption{The energies and transition matrix element for the first five Bloch states at $k_0=0$ of the 1D solid. For the four-level system, we use the Bloch state from level 1 to level 4 (see text).}
\label{tab:five_level_system_parameters}
\end{table}

\begin{figure*}[htb!]
\centering
\includegraphics[width=0.8\textwidth]{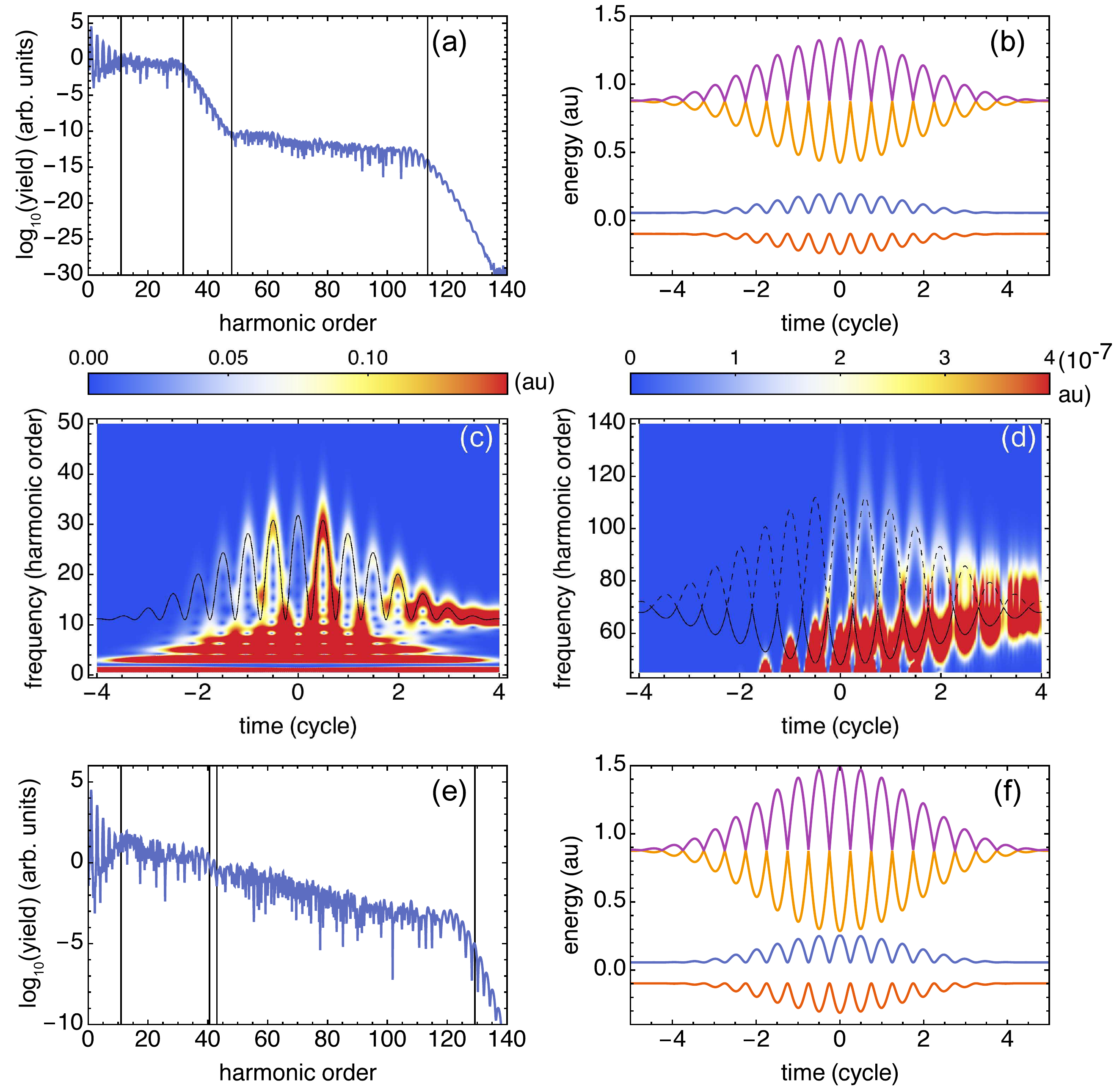}
\caption{(Color online) (a) HHG spectrum from a four-level system, with the field strength $E_0=4.1\times 10^{-3}$ au. The harmonic spectrum has two plateaus. The solid lines indicate the predictions from Eq.~(\ref{eqn:four_level_first_cutoff}-\ref{eqn:four_level_second_cutoff}) for the lowest and cutoff energies of the two plateaus. (b) The energy of the adiabatic states as a function of time. (c) The wavelet transform of time-dependent current in the energy range of the first plateau. The solid line overlaid on top is the energy difference between the first and the second adiabatic states. (d) The same as (c) but in the range of the second plateau. The solid line indicates the energy difference between the first and third adiabatic states, whereas the dashed line indicates the energy difference between the first and fourth adiabatic states. (e,f) The same as (a,b) but with a field strength $E_0=5.5\times 10^{-3}$ au. At this field strength, the second plateau merges onto the first plateau.} 
\label{fig:multilevel_adiabatic_states_and_spectrum}
\end{figure*}

The harmonic spectrum from the four-level system, driven by a 3.2~$\mu$m laser with a field strength of $E_0=4.1\times 10^{-3}$ au is shown in Fig.~\ref{fig:multilevel_adiabatic_states_and_spectrum}(a). We note that retaining more Bloch states in the TDSE calculation does not change the harmonic spectrum for a range of field strengths, see \cite{Wu2015}. The spectrum exhibits a characteristic two-plateau structure, where each plateau has a onset energy and a cutoff. The mechanism for this two-plateau structure can be understood by extending the three-step picture discussed for the two-level system in the previous section. The time-dependent energies of the adiabatic states for the four-level system are plotted in Fig.~\ref{fig:multilevel_adiabatic_states_and_spectrum}(b). The adiabatic states form two pairs of two closely coupled  states, resembling the adiabatic states of our two-level system in Fig.~\ref{fig:two_level_HHG}(b). The dynamics in the four-level system can then be understood similarly to that in the two-level system, where harmonics are generated by the transitions between the adiabatic states, except now we can have transitions from all three higher adiabatic states to the lowest adiabatic state. 

\begin{figure}[h]
\centering
\includegraphics[width=0.48\textwidth]{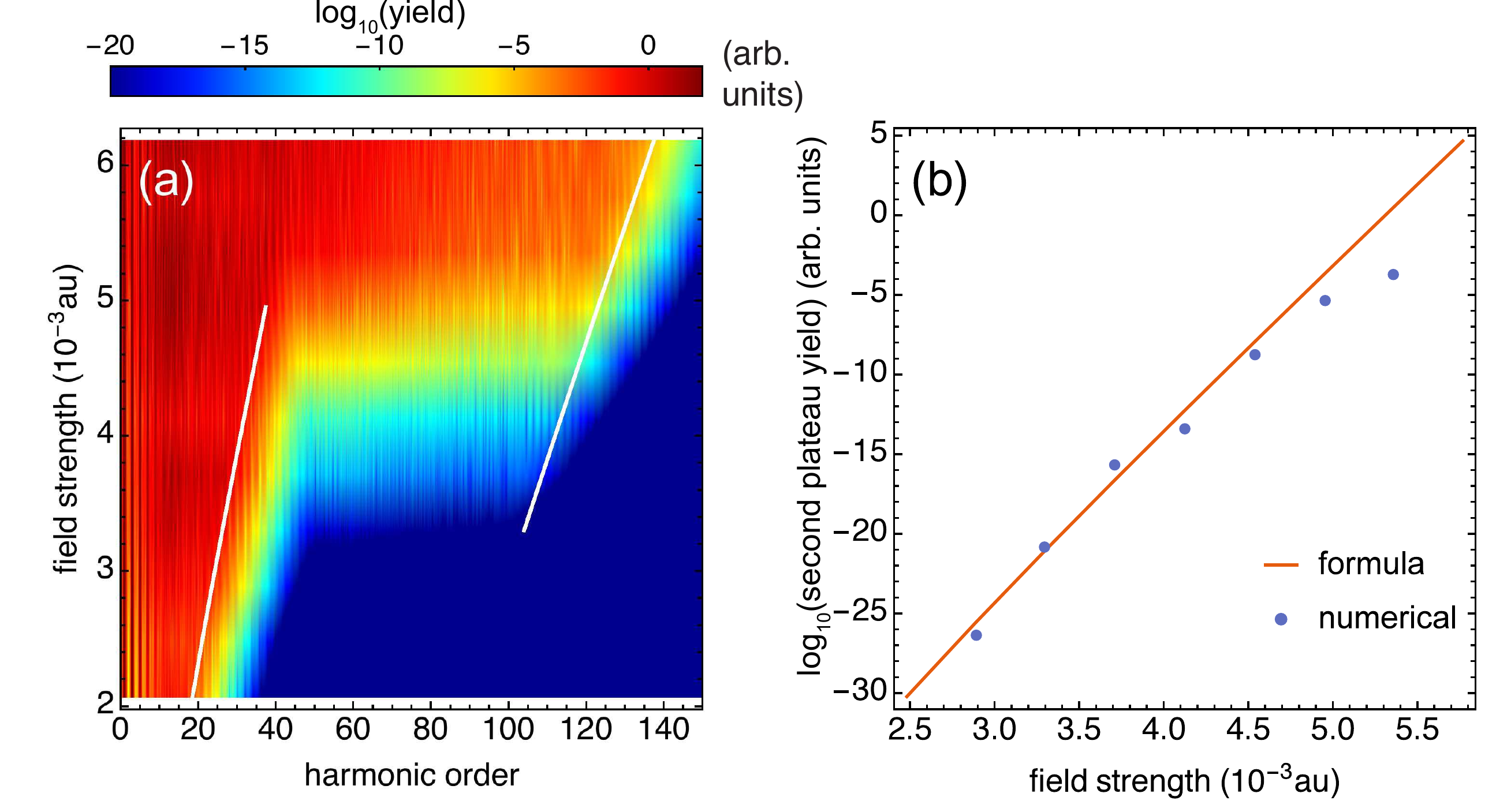}
\caption{(Color online) (a) Harmonic spectrum for a four-level system as a function of the field strength. The cutoffs predicted from Eq.~\eqref{eqn:four_level_first_cutoff} and Eq.~\eqref{eqn:four_level_second_cutoff} are indicated with white lines (b) The intensity of the second plateau (blue dots) agrees very well with the prediction from Eq.~\eqref{eqn:tunneling_rate_second_plateau} (red solid line). } 
\label{fig:four_level_HHG_intensity_scan}
\end{figure}

Fig.~\ref{fig:multilevel_adiabatic_states_and_spectrum}(c) shows the wavelet transform of the time-dependent current in the energy range of the first plateau, which describes the emission time of the harmonics in the first plateau. The black line overlaid on top is the energy difference between the first and second adiabatic states $\mathcal{E}_2(t)-\mathcal{E}_1(t)$. Fig.~\ref{fig:multilevel_adiabatic_states_and_spectrum}(d) shows the same wavelet transform but in the second plateau region. The dashed and solid black lines are the energy differences between the third, fourth and first adiabatic states, respectively. We can see that the adiabatic energies agree very well with the harmonic emission times in both the first and second plateau, suggesting that this adiabatic state description of the four-level dynamics works very well. Similarly to the two-level system, the allowed harmonics are then determined by the allowed energy range of the adiabatic states. We defined the energies of the adiabatic states at the peak of the vector potential as $\tilde{\mathcal{E}}_n$, where $n$ goes from 1 to 4. Then the first plateau is due to the transition between the first and the second adiabatic states, which has an start and cutoff of
\begin{align}
E^{(1)}_\text{low}&=\omega_2-\omega_1\\
E^{(1)}_\text{cutoff}&=\tilde{\mathcal{E}}_2-\tilde{\mathcal{E}}_1.
\label{eqn:four_level_first_cutoff}
\end{align}
Similarly, the second plateau is due to the transition from the third and fourth adiabatic state to the first adiabatic states, which has an start and cutoff
\begin{align}
E^{(2)}_\text{low}&=\tilde{\mathcal{E}}_3-\tilde{\mathcal{E}_1}\\
E^{(2)}_\text{cutoff}&=\tilde{\mathcal{E}_4}-\tilde{\mathcal{E}_1}.
\label{eqn:four_level_second_cutoff}
\end{align}
These four energies are indicated by the vertical lines in Fig.~\ref{fig:multilevel_adiabatic_states_and_spectrum}(a). We can see that the beginning and end of the two plateaus from these formula agree very well with the plateau spans that can be seen in the harmonic spectrum. The agreement between the predicted cutoff energies and the numerical results is very good for a range of intensities, as shown in the intensity scan in Fig.~\ref{fig:four_level_HHG_intensity_scan}(a).

We turn next to the relative strength of the two harmonic plateaus in the four-level system. Similar to the two-level system, the intensity of the harmonic plateaus in the four-level system are determined by the population of the adiabatic states, which in turn are determined by the tunneling rates between them.
As shown in Tab.~\ref{tab:five_level_system_parameters}, the 1-2 and 3-4 couplings are much stronger than the 2-3 coupling, which means that the four-level system can be thought of as two weakly-coupled two-level systems. This means the intensity of the first plateau in the four-level system will be proportional to the tunneling rate $P_{12}$ between the first and second adiabatic states, given by the Landau-Zener rate (see Eq.~\ref{eqn:Landau-Zener_tunneling_rate})
\begin{equation}
I_{\rm 1st} \propto P_{12} = \exp \left(-\frac{\pi  (\omega_2-\omega_1)^2}{4 \omega  \mu_{12} A_0}\right).
\end{equation}

The intensity of the second plateau is more complicated. Since the four levels are coupled in a stepwise manner,  population transfer between the adiabatic states also happens in a stepwise manner, and the population on the fourth adiabatic state results from stepwise tunneling from state 1 to 2, 2 to 3, and 3 to 4. The total tunneling rate between 1 and 4 can be written as a product of these three tunnelings, 
\begin{equation}
I_{\rm 2nd} \propto P_{12}\times P_{23} \times P_{34}.
\label{2nd_plateau_eq}
\end{equation}
Since the two two-level systems are almost independent, the tunneling rates $P_{12}$ and $P_{34}$ are of the Landau-Zener form.
The 2-3 tunneling process is different because levels 2 and 3 are only weakly coupled to each other and there is no avoided crossing between them for low field strengths. The 2-3 tunneling will take place predominantly when the two levels have been pushed as close together as possible due to their respective interactions with levels 1 and 4, which happens at the peaks of the vector potential.
This tunneling event is therefore not field-driven in the same way that the Landau-Zener tunneling is. Instead, we treat it as simple tunneling in which the transition probability decreases exponentially with the barrier energy \cite{Ashcroft1976}:
\begin{equation}
P_{23} = \exp \left[-\beta  \left(\tilde{\mathcal{E}}_3-\tilde{\mathcal{E}}_2\right)\right],
\end{equation}
where $\tilde{\mathcal{E}}_3-\tilde{\mathcal{E}}_2$ is the energy difference between the two adiabatic states at the peak of the vector potential and $\beta$ is a constant. This minimum energy difference can be approximated as
\begin{equation}
\tilde{\mathcal{E}_3}-\tilde{\mathcal{E}}_2 \approx (\omega _3-\omega _2)- \left(\mu _{12}+\mu _{34}\right)A_0, 
\end{equation}
corresponding to the field free energy difference, minus the amount the two levels have been pushed together by their respective interactions with levels 1 and 4. 

Our analysis of the stepwise tunneling process that populates the fourth adiabatic state leads to an intensity of the second plateau that according to Eq.~\eqref{2nd_plateau_eq} is proportional to 
\begin{align}
I_{\rm 2nd} & \propto \exp \left[-\frac{\pi  \left(\omega _2-\omega _1\right){}^2}{4 \omega \mu _{12} A_0 }\right] \nonumber\\
&\times \exp \left[-\beta  \Big(\omega _3-\omega _2- \left(\mu _{12}+\mu _{34}\right)A_0\Big)\right] \nonumber\\
&\times \exp \left[-\frac{\pi  \left(\omega _4-\omega _3\right){}^2}{4 \omega \mu _{34} A_0}\right].
\label{eqn:tunneling_rate_second_plateau}
\end{align}
Fig.~\ref{fig:four_level_HHG_intensity_scan}(b) shows the comparison between the numerically calculated harmonic intensity of the second plateau and that from Eq.~\eqref{eqn:tunneling_rate_second_plateau} with the scaling parameter $\beta = 140$. We see they agree very well. At field strengths $E_0\geq 5.5\times 10^{-3}$ au, the two results start to deviate. This is because at fields higher than this, the adiabatic states 2 and 3 will cross each other in energy, and a simple exponential tunneling rate will break down. Furthermore, when $\tilde{\mathcal{E}}_2\approx\tilde{\mathcal{E}_3}$ the population very easily tunnels between the second and third adiabatic states and as a consequence, the second plateau will rise to a point where it almost merges with the first plateau. This can be  seen in the harmonic spectrum shown in  
Fig.~\ref{fig:multilevel_adiabatic_states_and_spectrum}(e), calculated at a field strength of $E_0=5.5\times 10^{-3}$ au. 

\begin{figure}[h]
\centering
\includegraphics[width=0.48\textwidth]{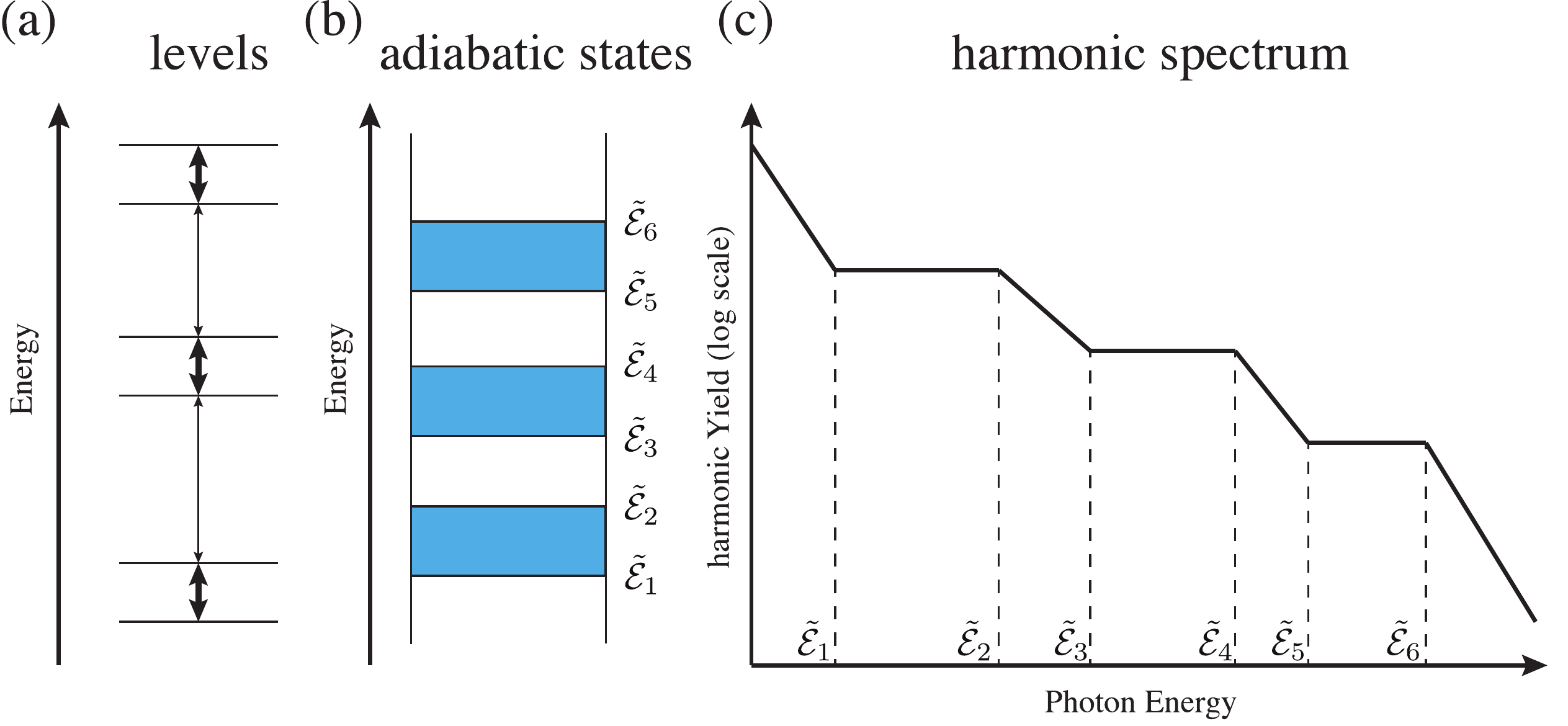}
\caption{(Color online)  (a) A schematic plot of a multi-level system where levels are coupled in pairs. (b) The energies of the adiabatic states form allowed regions indicated by the blue area. (c) Schematic plot of the multi-plateau structure from this multi-level system. The start and the cutoff of the plateau is determined by the bounds of the allowed energy regions.} 
\label{fig:schematic_HHG_multilevel_system}
\end{figure}

In the four-level system considered above, the harmonics are generated from two pairs of coupled adiabatic states. This picture of high harmonic generation can be generalized to systems with more than four levels, as illustrated in Fig.~\ref{fig:schematic_HHG_multilevel_system}. 
Fig.~\ref{fig:schematic_HHG_multilevel_system}(a) shows a multi-level system in which the levels are coupled in a stepwise manner. In particular, the levels in the multi-level system form pairs of strongly coupled two-level systems. The adiabatic states of this chain of multi-level system then form blocks of allowed energies, as indicated by the blue region in Fig.~\ref{fig:schematic_HHG_multilevel_system}(b). The harmonic spectrum generated by this multi-level system has multiple plateaus, as shown in Fig.~\ref{fig:schematic_HHG_multilevel_system}(c). The beginning and end of each plateau corresponds to the lower and higher bounds for the corresponding block of the adiabatic energies. More generally, if the multi-level system is not strongly coupled in pairs of states as we have considered here, the harmonic spectrum can exhibit even more plateaus. As an example, in the four-level system considered in Fig.~\ref{fig:multilevel_adiabatic_states_and_spectrum} levels three and four are near-degenerate and very strongly coupled to each other. A four-level system in which these levels were less degenerate would lead to harmonic spectra with three plateaus, corresponding to transitions from the second, third, and fourth adiabatic level to the lowest level, with cutoff energies each determined by the maximum adiabatic state energy separations.

\begin{figure}[h]
\centering
\includegraphics[width=0.4\textwidth]{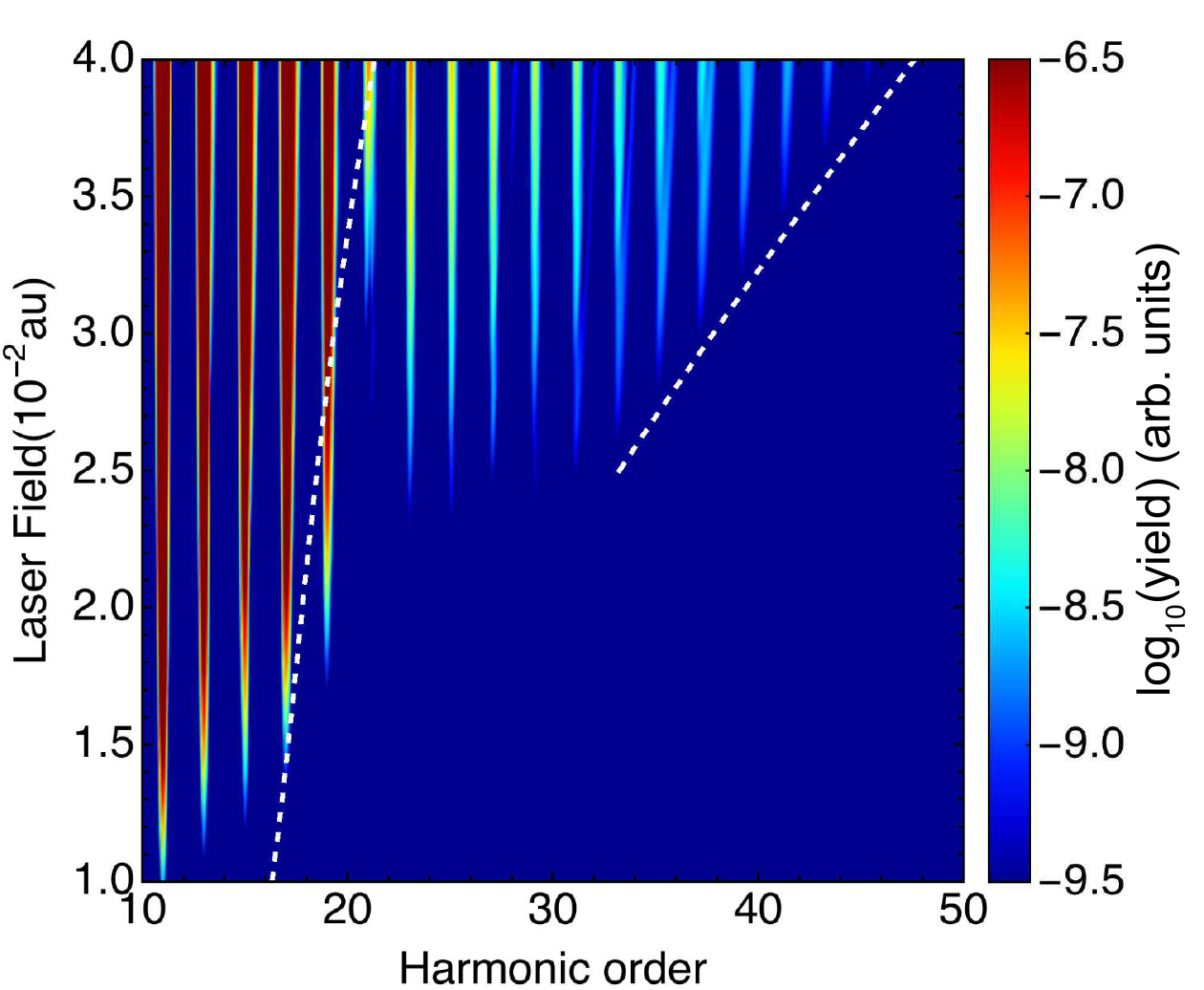}
\caption{(Color online) Intensity dependent harmonic spectra for the four-level model for solid argon. The dashed lines indicate the cutoff prediction from Eq.~\eqref{eqn:four_level_first_cutoff} and Eq.~\eqref{eqn:four_level_second_cutoff}.} 
\label{fig:solid_argon_HHG}
\end{figure}

Finally, we end this section by returning to the experimental observations of multiple plateaus in solid argon and krypton as reported in \cite{NdabashimiyeG.2016a}. The calculations describing HHG in solid argon in \cite{NdabashimiyeG.2016a} were based on a multi-level model similar to that described above, except the energy levels and transition matrix elements originated in a density functional theory (DFT) calculation of the 3D argon band structure. The energies of the four lowest bands at the $\Gamma$ point in solid argon are 0, 14 eV, 20 eV, and 29 eV, and we found that it was sufficient to include these four levels to get converged harmonic spectra. 
We used the experimental observations of the cutoff energies and relative strengths of the first and second plateau to adjust the transition matrix elements between the different levels from the DFT predictions. In particular, the experiment observed a rapid increase of the secondary plateau with intensity, so that it nearly matched the strength of the primary plateau, similar to the situation discussed in the context of Fig.~\ref{fig:multilevel_adiabatic_states_and_spectrum}(e). Also, the experiment observed that the difference between the two cutoff energies 
(approximately 25 and 33 eV, respectively, at an intensity of 20 TW/cm$^2$ and wavelength of 1333 nm) was relatively modest given the separation between levels 2, 3, and 4. We found the best agreement with the experimental results when using step-wise coupling matrix elements of approximately equal strength (as opposed to the DFT prediction of a weak coupling between levels 2 and 3) 
so that the relatively strong coupling between levels two and three acts to both cap the cutoff energy of the primary plateau, and gives rise to a strong secondary plateau as discussed above. The resulting intensity dependent spectra are shown in Fig.~\ref{fig:solid_argon_HHG}. We note that although the third plateau, due to the transitions between levels four and one, is so weak it is not visible on the scale shown in the figure, the fourth level plays an important role in capping the cutoff energy of the secondary plateau. 

\section{Semi-classical picture of HHG in solid}
\label{sec:Semiclassical_picture}
In Section ~\ref{sec:TDSEsolid} we showed that the electron dynamics in solids can be formulated as dynamics in a multi-level system, and we have provided a three-step picture in Sections ~\ref{sec:two_level_HHG} and \ref{sec:HHG_multi-level_system} for harmonic generation in the multi-level system. We will now come back to the study of the connection between a model solid and a multi-level system. We will show that this connection provides a semi-classical picture for the harmonic generation process by delocalized electrons in a solid. In Appendix C we will show that this picture can also be arrived at via applying the strong field approximation (SFA) \cite{Lewenstein1994,Lewenstein1995,Brabec2000,Milosevic2006,Krausz2009} to the dynamics of strongly driven Bloch electrons. 

\begin{figure}[h]
\centering
\includegraphics[width=0.48\textwidth]{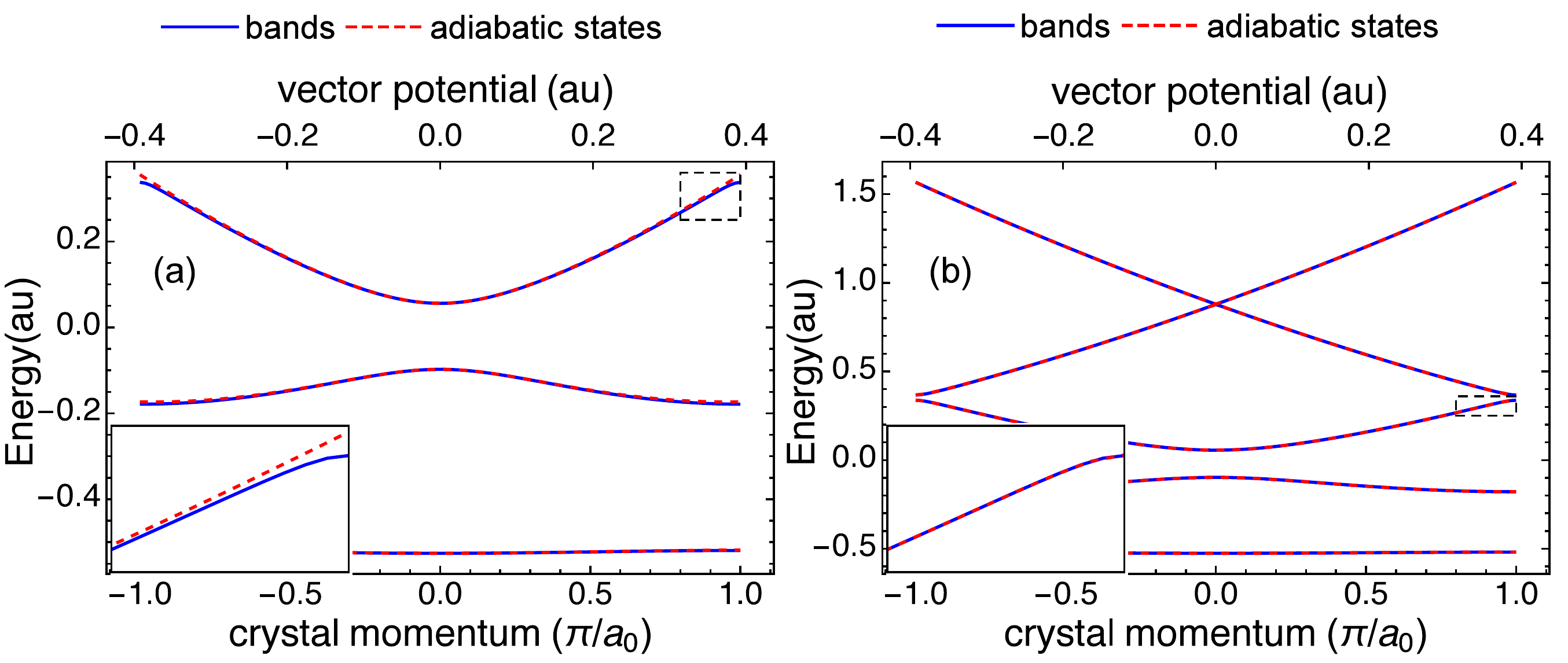}
\caption{(Color online) Comparison of the band structure (blue solid lines, bottom axis) and the energy of the adiabatic states vs vector potential (red dashed lines, top axis) for (a) a three-level and (b) a five-level system. The insets in both (a) and (b) show a zoom-in of the region near the edge of the third band, indicated by boxes.} 
\label{fig:bands_VS_adiabatic_states}
\end{figure}

We start by considering the evolution of the adiabatic state energies with field strength. As discussed in Section~\ref{sec:HHG_multi-level_system} the maximum difference between these energies at a given field strength will determine the cutoff energies of the different plateaus. Fig.~\ref{fig:bands_VS_adiabatic_states}(a) shows the energies of the adiabatic states for a three-level system as functions of driving vector potential (red dashed lines, vector potential along top axis). We compare these to the band structure of the three lowest bands in our model solid, {\it i.e.}, we plot the energies of the bands as functions of crystal momentum (blue solid lines, crystal momentum along the bottom axis). We see that the energies of the adiabatic states, when driven by a laser field with a vector potential $A_0$, map out the band structure very well when we assign a value of $k=A_0$ to the crystal momentum. 

The origin of this agreement can be understood by considering the Houston state basis for the solid, which is related to the Bloch state basis via a unitary transformation (see \cite{Wu2015} for details). The Houston states are constructed as the instantaneous eigenstates of the time-dependent Hamiltonian, which includes both the crystal potential and the field in the dipole approximation. The Houston states are thus the adiabatic states by definition. They are characterized by a time-dependent momentum $k(t) = k_0 + A(t)$, where $k_0$ is the value of the crystal momentum in the absence of the field (corresponding to the Bloch state with $k=k_0$). 
The energies of the Houston states $\tilde{\varepsilon}_{n}(t)$ are simply the energies of the Bloch states at the crystal momentum that corresponds to the vector potential \cite{Krieger1986,Wu2015}:
\begin{equation}
\tilde{\varepsilon}_{n}(t) = \varepsilon_{n}(k(t)) - \frac{1}{2}A^2(t),~{\rm with}~k(t) = k_0+A(t),
    \label{eqn:adibatic_energy_band_relationship}
\end{equation}
where the constant term comes from the fact that Bloch and Houston Hamiltonians differ by a $\frac{1}{2}A^2$ term (see \cite{Wu2015}). 

An important implication of Eq.~(\ref{eqn:adibatic_energy_band_relationship}) is that the band structure at all $k$ is encoded at $k_0=0$ (or any other initial $k_0$) through the energies of the Bloch states and transition matrix elements between them. This means that as long as our multi-level system is constructed from the energies and coupling strengths of our model solid, its evolution with vector potential is in fact equivalent to the evolution of the Houston states and should indeed yield the band structure. In fact, the Houston description is the adiabatic state description of the dynamics in the Bloch state basis. A complete proof of this relationship can be found in Appendix B. 

We note that the agreement between the adiabatic states and the band structure improves with the inclusion of more levels in the multi-level model. This is illustrated in the comparison between Fig.~\ref{fig:bands_VS_adiabatic_states}(a) and (b), in particular as magnified in the insets. In a multi-level system, the energies of the adiabatic states are determined both by their interactions with the laser field and with each other.
The former leads to an approximately linear variation with field strength whereas the latter leads to avoided crossings. In a multi-level description, it is the avoided crossings with higher-lying levels that allow lower-lying levels to ``turn over" at high field strengths and thus mimic the band structure at the edge of the Brillouin zone, as seen in the inset in Fig.~\ref{fig:bands_VS_adiabatic_states}(b). In contrast, in the absence of higher-lying levels, the adiabatic energy will continue to increase linearly as seen in the  Fig.~\ref{fig:bands_VS_adiabatic_states}(a) inset.

\begin{figure}[t]
\centering
\includegraphics[width=0.35\textwidth]{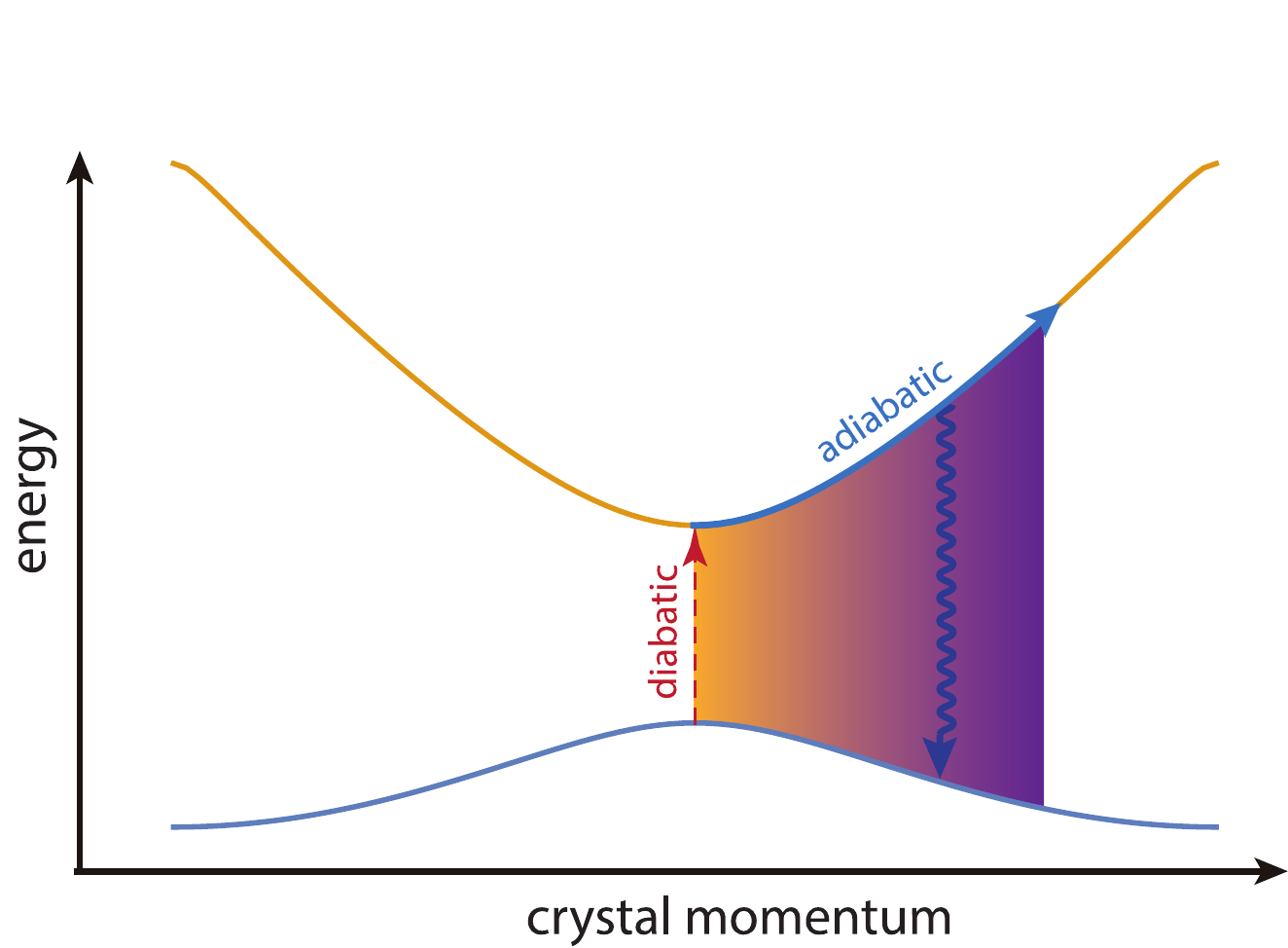}
\caption{(Color online) A schemetic plot of the momentum space three-step picture for harmonic generation in solids. In the first step, the valence electron tunnels through the band gap. In the second step, the electron wave function evolves adiabatically in the valence and conduction bands. In the final step, the coherence between the  valence and conduction band states leads to emission of radiation with energies corresponding to the instantaneous energy difference between the dressed valence and conduction bands.} 
\label{fig:three_step_scheme}
\end{figure}

Another important implication of the result in Fig.~\ref{fig:bands_VS_adiabatic_states} and the energy relationship described in  Eq.~(\ref{eqn:adibatic_energy_band_relationship}) is that it offers a picture of how the harmonic generation process for Bloch electrons in solids takes place in three steps, similarly to the well known three-step model for harmonic generation in gases. This picture is illustrated in Fig.~\ref{fig:three_step_scheme} for a two-band system. In the first step, which in time corresponds to the zeroes of the vector potential, a part of the delocalized electron wavepacket diabatically tunnels through the band gap into the conduction band via Landau-Zener tunneling. In the second step, the valence and conduction band wave functions both evolve adiabatically, accumulating phases according to their instantaneous energies. In the third step, the coherence between the conduction band and the valence electron leads to the emission of radiation with an energy corresponding to the instantaneous band gap. The third step leads to emission during the entire half-optical-cycle, from one zero of the vector potential to the next, where each energy below the cutoff is emitted at two different times as the vector potential increases and decreases, as is also shown in  Fig.~\ref{fig:multilevel_adiabatic_states_and_spectrum}(c). The cutoff energy is determined by the instantaneous band gap at the peak of the vector potential:
\begin{equation}
    E_\text{cutoff} = \varepsilon_c(A_0) - \varepsilon_v(A_0),
\end{equation}
and can therefore not exceed the maximum band gap of the two-band system. In a multi-band system, this generalizes to the conclusion that the cutoff energies of each plateau will be limited by the maximum separation between the valence band and the upper-most conduction band responsible for that plateau, 
\begin{equation}
    E^n_\text{cutoff} = \varepsilon_{nc}(A_0) - \varepsilon_v(A_0),
    \label{eqn:cutoff_multiple_bands}
\end{equation}
as illustrated in Fig.~\ref{fig:schematic_HHG_multilevel_system}. 

The cutoff formula above, and its limit of the maximum band gap in the case of a two-band system, recovers the cutoff formula discussed by Vampa and collaborators in \cite{Vampa2015}. In \cite{Vampa2015}, a {\it real space} three-step picture of HHG in solids is discussed in terms of localized electron and hole separation, acceleration and recollision. The laser pulse generates electron-hole pairs and they travel in opposite directions in space and then recombine as the vector potential changes sign. A similar real space picture was also presented by Higuchi {\it et al.} in \cite{Higuchi2014a}, where the dynamics is described in terms of (localized) Wannier states and the recollision is described in terms of the electron recombination to  neighboring cores. In contrast, in this paper we are discussing a {\it momentum space} three-step picture of the HHG process, in terms of adiabatic and diabatic evolution of the initial (delocalized) Bloch state. The harmonics in this picture comes from the coherence between the one-electron valence and conduction band Bloch states.

\section{Summary}
\label{sec:Summary}

In this paper, we have investigated HHG in a multi-level system and discussed how such a model can be used to describe HHG in bulk solids, a subject that is currently of great interest in the ultrafast community \cite{Ghimire2010, Ghimire2011, Schubert2014, Hohenleutner2015, Mahmood2015,Vampa2015a, NdabashimiyeG.2016a}. We have shown that a driven multi-level system generally gives rise to harmonic spectra exhibiting multiple plateaus, and that the cutoff energies and relative strengths of each plateau can be calculated using simple formulas. We discussed how the strong-field dynamics is best described in the adiabatic (dressed) state basis for the multi-level system. In this basis, the HHG process happens by tunneling of the population from a lower to an upper state, followed by evolution of the population on the excited state and finally transition back to the ground state. This means that the cutoff energies of each plateau at a given field strength are simply determined by the maximum energy difference between the field-dressed adiabatic upper and lower states. The strength of each plateau, and in particular its dependence on field strength, can be calculated considering a sequence of tunneling events. In a four-level system in which the states are strongly coupled in pairs, the secondary plateau is due to transitions from the third and fourth adiabatic states to the first adiabatic states, where the highest state is reached via tunneling from 1-2, 2-3, and finally 3-4. We also discussed how a detailed knowledge of driven multi-level dynamics can be used in reverse to draw conclusions from experimental HHG results. We used the recent example of HHG in solid argon, in which the appearance of multiple plateaus and their relative extents and strengths was used as an indication of the contributions and relative couplings of several different conduction bands \cite{NdabashimiyeG.2016a}. 

We also showed that if the multi-level system originates in the $k=0$ component of the band structure of a periodic system, the adiabatic states of the multi-level system will map out the entire band structure as the field strength is increased. This means that the maximum cutoff energies of each plateau will be limited by the band structure - the highest photon energy emitted in a given plateau will be limited by the largest energy separation in the Brillouin zone between the valence band and the conduction band which is responsible for that particular plateau. This could potentially be used to map out the band structure, including the high-lying bands in the solid, by careful measurements of how different cutoff energies scale with intensity, in analogy with our discussion of solid argon above. For materials that have very different band structure along different crystal orientations, one can also expect to see the crystal orientation dependent harmonics and mapping out the 3D band structure by measurements of how different cutoff energies scale with intensity along different crystal orientation. 

The correspondence between the adiabatic states and the band structure also leads naturally to a semi-classical, three-step picture for HHG by Bloch electrons in a solid: In the first step, a part of the delocalized electron wavepacket diabatically tunnels through the band gap into the conduction band via Landau-Zener tunneling. In the second step, the valence and conduction band wave functions both evolve adiabatically, accumulating phases according to their instantaneous energies. In the third step, the coherence between the conduction band and the valence electron leads to the emission of radiation with an energy corresponding to the instantaneous band gap. The harmonic emission is thus chirped in time, at the sub-cycle level, similarly to what happens in gas-phase HHG. 

This momentum space picture can potentially also shed light on the measured ellipticity dependence of harmonics in solid argon as reported in \cite{NdabashimiyeG.2016a}, where the yield of the harmonics in the plateau was found to decrease exponentially with ellipticity similarly to what is found in gas-phase argon. In a linearly polarized field, the valence to conduction band tunneling of the electron takes place twice each laser cycle around the $\Gamma$ point when the electron traverses the Brillouin zone center. The intensity of the harmonics in the plateau is then determined by this tunneling rate around the $\Gamma$ point. For a circularly polarized field, the electron will rotate around the $\Gamma$ point in momentum space, without ever getting close to the $\Gamma$ point, thus the tunneling rate is greatly suppressed. 
The details of the ellipticity dependence of HHG in solids will be studied in a future paper.

\section{Acknowledgement}

We acknowledge valuable discussions with the experimental group of David Reis and Shambhu Ghimire, and with Dana Browne and Fran\c{c}ois Mauger at LSU. This work was supported by the National Science Foundation under Grant No. PHY-1403236. High-performance computational resources were provided by Louisiana Optical Network Initiative (www.loni.org) and by the High Performance Computer Centre at LSU (www.hpc.lsu.edu).

\appendix
\begin{widetext}
\section{Adiabatic basis formalism}

In this appendix we describe the formalism for a multi-level system in the adiabatic basis. The time-dependent Schr\"odinger equation reads
\begin{equation}
	i \ket{\dot\psi(t)} = H(t) \ket{\psi(t)},
\end{equation}
and the Hamiltonian is
\begin{align}
H(t) &= H_0 + A(t)\hat{p},
\label{eqn:adiabatic_general_H}
\end{align}
where $H_0$ is the laser free Hamiltonian, $A(t)$ is the vector potential, and $\hat{p}$ is the dipole operator in momentum space.
The adiabatic states are the instantaneous eigenstates of the time-dependent Hamiltonian
\begin{equation}
	H(t)\ket{\phi_n(t)} = \mathcal{E}_n(t) \ket{\phi_n(t)},
\label{eqn:eigen_equation_for_adiabatic_states}
\end{equation}
where $\ket{\phi_n(t)}$ is an adiabatic state and $\mathcal{E}_n(t)$ is its energy. Since the adiabatic states are the instantaneous eigenstates of the time-dependent Hamiltonian, they are time-dependent themselves. However, at any given time $t$ they are orthogonal and form a complete basis set:
\begin{equation}
\braket{\phi_n(t)|\phi_m(t)} = \delta_{nm}.
\end{equation}
This means that the time-dependent wave function $\ket{\psi(t)}$ can be expanded in the adiabatic basis as
\begin{equation}
	\ket{\psi(t)} = \sum_n \tilde{C}_n(t)\ket{\phi_n(t)}.
	\label{eqn:multilevel_wfn_in_adiabatic_basis}
\end{equation}
where $\tilde{C}_n(t)$ is the amplitude of the adiabatic state $\ket{\phi_n(t)}$. In the following, we will leave out the explicit indication of time-dependence and write $\ket{\phi(t)}$ and $\tilde{C}_n(t)$ as $\ket{\phi}$ and $\tilde{C}_n$, respectively, for brevity. Substituting the wave function into the Schr\"odinger equation, we have
\begin{equation}
	i \sum_n \dot{\tilde{C}}_n \ket{\phi_n} + i \sum_n \tilde{C}_n\ket{\dot{\phi_n}} = \sum_n \tilde{C}_n H\ket{\phi_n}.
\end{equation}
Projecting onto one of the adiabatic states, we obtain the following equation:
\begin{equation}
	i \dot{\tilde{C}}_n  + i \sum_m \tilde{C}_m \braket{\phi_n|\dot{\phi_m}} = \tilde{C}_n \mathcal{E}_n.
\label{eqn:eqn_for_adiabatic_coefficients_1}
\end{equation}
To calculate the term $\braket{\phi_n|\dot{\phi_m}}$, we take the derivative of Eq.~\eqref{eqn:eigen_equation_for_adiabatic_states} and project onto one of the adiabatic states:
\begin{align}
	\braket{\phi_n|\dot H|\phi_m} + \mathcal{E}_n \braket{\phi_n|\dot{\phi_m}} = \mathcal{\dot{E}}_m \delta_{mn} + \mathcal{E}_m \braket{\phi_n|\dot{\phi_m}}.
\end{align}
This gives
\begin{equation}
	\begin{array}{cc}
		\braket{\phi_n|\dot{\phi}_m} = \left\{ 
		\begin{array}{cc}
			E(t)\frac{\braket{\phi_n|\hat{p}|\phi_m}}{\mathcal{E}_n-\mathcal{E}_m} & m\neq n \\[1em]
			0 & m=n \\
		\end{array}
		\right.
		\\
	\end{array},
\label{eqn:phi_phi_dot}
\end{equation}
where $E(t) = -\dot{A}(t)$ is the electric field. Substituting Eq.~\eqref{eqn:phi_phi_dot} and Eq. \eqref{eqn:adiabatic_general_H}  into Eq. \eqref{eqn:eqn_for_adiabatic_coefficients_1},  we finally obtain the TDSE in the adiabatic basis
\begin{equation}
i  \dot{\tilde{C}}_n(t) =  \sum_m \left[\mathcal{E}_m(t) \delta_{mn} - E(t) X_{mn}(t)\right] \tilde{C}_m(t),
\label{eqn:EOM_two_level_adiabatic_state}
\end{equation}
where
\begin{equation}
\begin{array}{cc}
X_{nm}(t) = \left\{ 
\begin{array}{cc}
 i\frac{\braket{\phi_n|\hat{p}|\phi_m}}{\mathcal{E}_n-\mathcal{E}_m} & m\neq n \\[1em]
0 & m=n \\
\end{array}
\right.
\\
\end{array}
\label{eqn:X_matrix_general_adiabatic_formalism}
\end{equation}
indicates the transition matrix element between the adiabatic states. For a two-level system driven by a continuous wave $A(t)=A_0 \sin\omega t$, the transition matrix element has the form
\begin{equation}
X_{12}(t) = \frac{i\omega_0\mu}{\omega^2_0+4A^2_0\mu^2 \sin^2\omega t},
\label{eqn:X_matrix_two-level}
\end{equation}
where $\mu$ is the matrix element between Bloch states (Eq.~4). For the strongly driven case such that $\omega_0/2A_0\mu \ll 1$, $X_{12}(t)$ is small except near the zeroes of the vector potential. This means that the population exchange between the adiabatic states  happens only around the avoided crossings. 

The time-dependent current in the adiabatic basis can be calculated as
\begin{align}
    j(t) = \braket{\psi|\hat p |\psi}.
\end{align}
Substituting Eq.~\eqref{eqn:multilevel_wfn_in_adiabatic_basis} into the above equation, we can separate the current into two parts
\begin{align}
    j(t) = j_\text{intra}(t) + j_\text{inter}(t),
\end{align}
where $j_{\text{intra}}$ and $j_{\text{inter}}$ are the contribution from the dynamics in adiabatic states and between adiabatic states.
\begin{align}
    j_{\text{intra}}(t) &= \sum_n \left|\tilde{C}_n \right|^2 \braket{\phi_n|\hat p|\phi_n},\\
    j_{\text{inter}}(t) &= \sum_{\substack{m,n \\ m \neq n}} \tilde{C}^*_m \tilde{C}_n\braket{\phi_m|\hat p|\phi_n}.
\end{align}
In the context of strong-field dynamics in a solid, describing the dynamics in terms of the adiabatic states is equivalent to describing it in terms of the well-known Houston states since these are in fact the adiabatic states of the laser-driven solid \cite{Wu2015}. The contributions from $j_{\text{intra}}(t)$ and $j_{\text{inter}}(t)$ are equivalent to the intra-band and inter-band contributions, respectively.

\section{Adiabatic states in a multi-level system: connection to band structure}

In this appendix, we show that the energies of the adiabatic states for the multi-level system formed by the Bloch states at $k_0=0$ trace out the band structure of the solid. 

The Bloch states are the eigenstates of the laser-free Hamiltonian
\begin{equation}
\left[\frac{\hat{p}^2}{2}+V(x)\right]\ket{\phi_{nk_0}} =  \varepsilon_n(k_0) \ket{\phi_{nk_0}}.
\label{eqn:TIDSE_Bloch_state_appendix}
\end{equation}
The adiabatic states for the multi-level system formed by the Bloch state at $k_0$ are the instantaneous eigenstates of the system when a vector potential $A$ is applied:
\begin{equation}
\left[\frac{\hat{p}^2}{2}+V(x) +A\hat{p}\right]\ket{\tilde\phi_{nk_0}} =  \mathcal{E}_n(A) \ket{\tilde\phi_{nk_0}},
\end{equation}
where $\mathcal{E}_n(A)$ is the instantaneous energy of the $n^{th}$ adiabatic state, and depends on instantaneous vector potential $A$. The above equation can be written as
\begin{equation}
\left[\frac{(\hat{p}+A)^2}{2}+V(x) - \frac{A^2}{2}\right]\ket{\tilde\phi_{nk_0}} =  \mathcal{E}_n(A) \ket{\tilde\phi_{nk_0}}.
\label{eqn:TDSE_adiabatic_appendix}
\end{equation}
The Hamiltonians in Eqs.~\eqref{eqn:TDSE_adiabatic_appendix} and \eqref{eqn:TIDSE_Bloch_state_appendix} differ by a constant. This means that the time-dependent adiabatic wave function is related to the stationary Bloch state wave function by a phase factor: 
\begin{align}
\ket{\tilde\phi_{nk_0}} = e^{iA^2 t/2}\ket{\phi_{nk_0+A}}.
\end{align}
The adiabatic state energies then can be written in terms of the band energies
\begin{equation}
\mathcal{E}_n(A) = \varepsilon_n(k_0) - \frac{A^2}{2},
\end{equation}
so the adiabatic state energies map out the band structure.

\section{Strong field approximation for solids}

In this appendix, we apply the strong field approximation (SFA) \cite{Lewenstein1994,Lewenstein1995,Milosevic2006} to study strong-field dynamics in a two-band solid using a Bloch state as the initial condition. We will show that the momentum-space three-step picture of HHG discussed in Section~\ref{sec:Semiclassical_picture} also naturally follows from such an SFA treatment. We start from the TDSE:
\begin{equation}
i\frac{\pd}{\pd t}\ket{\psi(t)} = \left[\frac{\left(\hat{p} + A(t)\right)^2}{2} +V(x)\right]\ket{\psi(t)}.
\label{eqn:Houston_TDSE}
\end{equation}
where $V(x)$ is the periodic potential and $A(t)$ is the vector potential. The  wave function can be expanded using the Houston states as
\begin{equation}
\ket{\psi(t)} = \sum_n a_{nk_0}(t) \ket{\widetilde\phi_{nk_0}(t)},
\label{eqn:wavefunction_expand_in_Houston_basis}
\end{equation}
where $\ket{\widetilde\phi_{nk_0}(t)}$ are the Houston states and $a_{nk_0}(t)$ the expansion coefficients. Substituting Eq.~\eqref{eqn:wavefunction_expand_in_Houston_basis} into Eq.~\eqref{eqn:Houston_TDSE}, the TDSE has the form \cite{Krieger1986,Wu2015}
\begin{equation}
i \frac{\partial a_{nk_0}}{\partial t} = \sum_{m}\Big[\varepsilon_n(k(t)) \delta_{nm} - E(t) X_{nm}(k(t)) \Big]a_{mk_0}.
\label{eqn:EOM_Houston}
\end{equation}
where $E(t) = -\dot{A}(t)$ is the electric field and $X_{nm}$ is the transition matrix elements between the Houston states. Note that  Eq.~\eqref{eqn:EOM_Houston} has the same form as Eq.~\eqref{eqn:EOM_two_level_adiabatic_state}, because the Houston states are the adiabatic states for the laser-dressed solid. For simplicity, we consider only the valence and conduction bands and limit ourselves to $k_0=0$, then the TDSE simplifies to
\begin{align}
i \frac{\partial a_{v}}{\partial t} &= \varepsilon_v(k(t)) a_{v}  - E(t) X_{vc}(k(t)) a_{c},\\
i \frac{\partial a_{c}}{\partial t} &= \varepsilon_c(k(t)) a_{c}  - E(t) X_{cv}(k(t)) a_{v},
\end{align}
where $k(t) = k_0 +  A(t)$.
In analogy with the treatment in the SFA for gas-phase HHG, we assume $\left|a_{v}\right|\approx 1$ so the valence and conduction band amplitudes can be integrated analytically:
\begin{align}
a_v (t) &=\exp\left[-i\int_0^t \varepsilon_v(k(t')) dt'\right],\\
a_c (t) &=i\int_0^t dt' \exp\left[-i\int_0^{t'} \varepsilon_v(k(t'')) dt''\right]\, E(t')\, X_{cv}(k(t'))\, \exp\left[- i\int_{t'}^t \varepsilon_c(k(t''))dt''\right].
\end{align}
The time-dependent current is then
\begin{align}
j(t)  &= a^*_v(t)a_{c}(t)X_{vc}\left[k(t)\right] + c.c.\nonumber\\
      &= i\int_0^t  X_{vc}(k(t))  e^{- i S(t,t')} E(t')\, X_{cv}(k(t')) dt'  + c.c.,
\label{eqn:SFA_dipole}
\end{align}
where the action is
\begin{equation}
S(t,t') = \int_{t'}^t \left[ \varepsilon_c(k(t'')) - \varepsilon_v(k(t'')) \right]dt''.
\end{equation}
The three terms in \eqref{eqn:SFA_dipole} correspond to recombination, propagation and ionization, respectively, and the harmonic spectrum is calculated as the Fourier transform of the time-dependent current. By requiring the action to be stationary, we get two saddle point conditions:
\begin{align}
\varepsilon_c(k(t')) - \varepsilon_v(k(t')) &= 0,\\
\varepsilon_c(k(t)) - \varepsilon_v(k(t))  &= E_r,
\end{align}
where $t'$ is the ionization time, $t$ is the recombination time, and $E_r$ is the recombination energy. We note that, compared to the gas-phase SFA, we have one less saddle point equation (addressing the recollision time). This is because we have used the delocalized Bloch state as our initial condition, which is coupled to only one of the Houston states when the laser is on. In contrast, in the gas-phase SFA the ground state is localized in space and couples to multiple Volkov states. 

The first saddle point equation gives us the ionization time, while the second saddle point equation gives the relation between the energy and the emission time of the emitted harmonic. For a simple vector potential $A=A_0\sin(\omega t)$, the time-dependent crystal momentum is $k(t) = A_0\sin\omega t$, and the solution to the first saddle point equation is:

\begin{equation}
t'= \frac{m\pi}{\omega} + i \gamma,~~~~~~~~~m=0,\pm 1, \pm 2 \cdots
\end{equation}
The real part of $t'$ is the ionization time and the imaginary part $\gamma$ is the valence-to-conduction band tunneling time, at the band gap. The actual form of $\gamma$ depends on the shape of the band structure around $k=0$. To lowest order in $k$, the band structure around $k=0$ can be approximated as
\begin{equation}
\varepsilon_c(k) - \varepsilon_v(k) \approx E_g + \frac{a}{2}k^2,
\end{equation}
where $E_g$ is the band gap and the $a$ is an expansion coefficient. Then the second saddle point equation becomes
\begin{equation}
E_g + \frac{a}{2} \left( A_0\sin(\omega t') \right)^2 = 0,
\end{equation}
which gives
\begin{equation}
t'=\frac{m\pi}{\omega} + \frac{i}{\omega} \sinh^{-1}\left(\sqrt{\frac{2E_g}{aA_0^2}}\right).
\end{equation}
This means that the tunneling mostly happens at the time when the vector potential is zero, and the electron is at the minimum band gap for the two bands.

The second saddle point equation then gives us the emission time for harmonics. For example, the energy of the emitted harmonic is maximized when the vector potential is at its maximum, and the recombination time for the cutoff harmonic is therefore
\begin{equation}
t= \frac{(2n+1)\pi}{\omega}.
\end{equation}
The maximum energy of the harmonic is then
\begin{equation}
E_r = \varepsilon_c(A_0) - \varepsilon_v(A_0),
\end{equation}
which is just the maximum bands the vector potential can sample. We note that the treatment presented here is similar to that in \cite{Vampa2015b}, but that here we stress a momentum-space three step picture of the HHG process.

\end{widetext}


\begin{thebibliography}{52}%
\makeatletter
\providecommand \@ifxundefined [1]{%
 \@ifx{#1\undefined}
}%
\providecommand \@ifnum [1]{%
 \ifnum #1\expandafter \@firstoftwo
 \else \expandafter \@secondoftwo
 \fi
}%
\providecommand \@ifx [1]{%
 \ifx #1\expandafter \@firstoftwo
 \else \expandafter \@secondoftwo
 \fi
}%
\providecommand \natexlab [1]{#1}%
\providecommand \enquote  [1]{``#1''}%
\providecommand \bibnamefont  [1]{#1}%
\providecommand \bibfnamefont [1]{#1}%
\providecommand \citenamefont [1]{#1}%
\providecommand \href@noop [0]{\@secondoftwo}%
\providecommand \href [0]{\begingroup \@sanitize@url \@href}%
\providecommand \@href[1]{\@@startlink{#1}\@@href}%
\providecommand \@@href[1]{\endgroup#1\@@endlink}%
\providecommand \@sanitize@url [0]{\catcode `\\12\catcode `\$12\catcode
  `\&12\catcode `\#12\catcode `\^12\catcode `\_12\catcode `\%12\relax}%
\providecommand \@@startlink[1]{}%
\providecommand \@@endlink[0]{}%
\providecommand \url  [0]{\begingroup\@sanitize@url \@url }%
\providecommand \@url [1]{\endgroup\@href {#1}{\urlprefix }}%
\providecommand \urlprefix  [0]{URL }%
\providecommand \Eprint [0]{\href }%
\providecommand \doibase [0]{http://dx.doi.org/}%
\providecommand \selectlanguage [0]{\@gobble}%
\providecommand \bibinfo  [0]{\@secondoftwo}%
\providecommand \bibfield  [0]{\@secondoftwo}%
\providecommand \translation [1]{[#1]}%
\providecommand \BibitemOpen [0]{}%
\providecommand \bibitemStop [0]{}%
\providecommand \bibitemNoStop [0]{.\EOS\space}%
\providecommand \EOS [0]{\spacefactor3000\relax}%
\providecommand \BibitemShut  [1]{\csname bibitem#1\endcsname}%
\let\auto@bib@innerbib\@empty
\bibitem [{\citenamefont {Ferray}\ \emph {et~al.}(1988)\citenamefont {Ferray},
  \citenamefont {L'Huillier}, \citenamefont {Li}, \citenamefont {Lompre},
  \citenamefont {Mainfray},\ and\ \citenamefont {Manus}}]{Ferray1988}%
  \BibitemOpen
  \bibfield  {author} {\bibinfo {author} {\bibfnamefont {M.}~\bibnamefont
  {Ferray}}, \bibinfo {author} {\bibfnamefont {A.}~\bibnamefont {L'Huillier}},
  \bibinfo {author} {\bibfnamefont {X.~F.}\ \bibnamefont {Li}}, \bibinfo
  {author} {\bibfnamefont {L.~A.}\ \bibnamefont {Lompre}}, \bibinfo {author}
  {\bibfnamefont {G.}~\bibnamefont {Mainfray}}, \ and\ \bibinfo {author}
  {\bibfnamefont {C.}~\bibnamefont {Manus}},\ }\href {\doibase
  10.1088/0953-4075/21/3/001} {\bibfield  {journal} {\bibinfo  {journal} {J.
  Phys. B At. Mol. Opt. Phys.}\ }\textbf {\bibinfo {volume} {21}},\ \bibinfo
  {pages} {L31} (\bibinfo {year} {1988})}\BibitemShut {NoStop}%
\bibitem [{\citenamefont {Farkas}\ \emph {et~al.}(1992)\citenamefont {Farkas},
  \citenamefont {T{\'{o}}th}, \citenamefont {Moustaizis}, \citenamefont
  {Papadogiannis},\ and\ \citenamefont {Fotakis}}]{Farkas1992}%
  \BibitemOpen
  \bibfield  {author} {\bibinfo {author} {\bibfnamefont {G.}~\bibnamefont
  {Farkas}}, \bibinfo {author} {\bibfnamefont {C.}~\bibnamefont {T{\'{o}}th}},
  \bibinfo {author} {\bibfnamefont {S.~D.}\ \bibnamefont {Moustaizis}},
  \bibinfo {author} {\bibfnamefont {N.~A.}\ \bibnamefont {Papadogiannis}}, \
  and\ \bibinfo {author} {\bibfnamefont {C.}~\bibnamefont {Fotakis}},\ }\href
  {\doibase 10.1103/PhysRevA.46.R3605} {\bibfield  {journal} {\bibinfo
  {journal} {Phys. Rev. A}\ }\textbf {\bibinfo {volume} {46}},\ \bibinfo
  {pages} {R3605} (\bibinfo {year} {1992})}\BibitemShut {NoStop}%
\bibitem [{\citenamefont {Harris}\ \emph {et~al.}(1993)\citenamefont {Harris},
  \citenamefont {Macklin},\ and\ \citenamefont {H{\"{a}}nsch}}]{Harris1993}%
  \BibitemOpen
  \bibfield  {author} {\bibinfo {author} {\bibfnamefont {S.}~\bibnamefont
  {Harris}}, \bibinfo {author} {\bibfnamefont {J.}~\bibnamefont {Macklin}}, \
  and\ \bibinfo {author} {\bibfnamefont {T.}~\bibnamefont {H{\"{a}}nsch}},\
  }\href {\doibase 10.1016/0030-4018(93)90250-9} {\bibfield  {journal}
  {\bibinfo  {journal} {Opt. Commun.}\ }\textbf {\bibinfo {volume} {100}},\
  \bibinfo {pages} {487} (\bibinfo {year} {1993})}\BibitemShut {NoStop}%
\bibitem [{\citenamefont {Schafer}\ \emph {et~al.}(1993)\citenamefont
  {Schafer}, \citenamefont {Yang}, \citenamefont {DiMauro},\ and\ \citenamefont
  {Kulander}}]{Schafer1993}%
  \BibitemOpen
  \bibfield  {author} {\bibinfo {author} {\bibfnamefont {K.~J.}\ \bibnamefont
  {Schafer}}, \bibinfo {author} {\bibfnamefont {B.}~\bibnamefont {Yang}},
  \bibinfo {author} {\bibfnamefont {L.~F.}\ \bibnamefont {DiMauro}}, \ and\
  \bibinfo {author} {\bibfnamefont {K.~C.}\ \bibnamefont {Kulander}},\ }\href
  {\doibase 10.1103/PhysRevLett.70.1599} {\bibfield  {journal} {\bibinfo
  {journal} {Phys. Rev. Lett.}\ }\textbf {\bibinfo {volume} {70}},\ \bibinfo
  {pages} {1599} (\bibinfo {year} {1993})}\BibitemShut {NoStop}%
\bibitem [{\citenamefont {Corkum}(1993)}]{Corkum1993}%
  \BibitemOpen
  \bibfield  {author} {\bibinfo {author} {\bibfnamefont {P.~B.}\ \bibnamefont
  {Corkum}},\ }\href {\doibase 10.1103/PhysRevLett.71.1994} {\bibfield
  {journal} {\bibinfo  {journal} {Phys. Rev. Lett.}\ }\textbf {\bibinfo
  {volume} {71}},\ \bibinfo {pages} {1994} (\bibinfo {year}
  {1993})}\BibitemShut {NoStop}%
\bibitem [{\citenamefont {Brabec}\ and\ \citenamefont
  {Krausz}(2000)}]{Brabec2000}%
  \BibitemOpen
  \bibfield  {author} {\bibinfo {author} {\bibfnamefont {T.}~\bibnamefont
  {Brabec}}\ and\ \bibinfo {author} {\bibfnamefont {F.}~\bibnamefont
  {Krausz}},\ }\href {http://rmp.aps.org/abstract/RMP/v72/i2/p545{\_}1}
  {\bibfield  {journal} {\bibinfo  {journal} {Rev. Mod. Phys.}\ }\textbf
  {\bibinfo {volume} {72}},\ \bibinfo {pages} {545} (\bibinfo {year}
  {2000})}\BibitemShut {NoStop}%
\bibitem [{\citenamefont {Sansone}\ \emph {et~al.}(2006)\citenamefont
  {Sansone}, \citenamefont {Benedetti}, \citenamefont {Calegari}, \citenamefont
  {Vozzi}, \citenamefont {Avaldi}, \citenamefont {Flammini}, \citenamefont
  {Poletto}, \citenamefont {Villoresi}, \citenamefont {Altucci}, \citenamefont
  {Velotta}, \citenamefont {Stagira}, \citenamefont {{De Silvestri}},\ and\
  \citenamefont {Nisoli}}]{Sansone2006}%
  \BibitemOpen
  \bibfield  {author} {\bibinfo {author} {\bibfnamefont {G.}~\bibnamefont
  {Sansone}}, \bibinfo {author} {\bibfnamefont {E.}~\bibnamefont {Benedetti}},
  \bibinfo {author} {\bibfnamefont {F.}~\bibnamefont {Calegari}}, \bibinfo
  {author} {\bibfnamefont {C.}~\bibnamefont {Vozzi}}, \bibinfo {author}
  {\bibfnamefont {L.}~\bibnamefont {Avaldi}}, \bibinfo {author} {\bibfnamefont
  {R.}~\bibnamefont {Flammini}}, \bibinfo {author} {\bibfnamefont
  {L.}~\bibnamefont {Poletto}}, \bibinfo {author} {\bibfnamefont
  {P.}~\bibnamefont {Villoresi}}, \bibinfo {author} {\bibfnamefont
  {C.}~\bibnamefont {Altucci}}, \bibinfo {author} {\bibfnamefont
  {R.}~\bibnamefont {Velotta}}, \bibinfo {author} {\bibfnamefont
  {S.}~\bibnamefont {Stagira}}, \bibinfo {author} {\bibfnamefont
  {S.}~\bibnamefont {{De Silvestri}}}, \ and\ \bibinfo {author} {\bibfnamefont
  {M.}~\bibnamefont {Nisoli}},\ }\href {\doibase 10.1126/science.1132838}
  {\bibfield  {journal} {\bibinfo  {journal} {Science}\ }\textbf {\bibinfo
  {volume} {314}},\ \bibinfo {pages} {443} (\bibinfo {year}
  {2006})}\BibitemShut {NoStop}%
\bibitem [{\citenamefont {Krausz}\ and\ \citenamefont
  {Ivanov}(2009)}]{Krausz2009}%
  \BibitemOpen
  \bibfield  {author} {\bibinfo {author} {\bibfnamefont {F.}~\bibnamefont
  {Krausz}}\ and\ \bibinfo {author} {\bibfnamefont {M.}~\bibnamefont
  {Ivanov}},\ }\href {\doibase 10.1103/RevModPhys.81.163} {\bibfield  {journal}
  {\bibinfo  {journal} {Rev. Mod. Phys.}\ }\textbf {\bibinfo {volume} {81}},\
  \bibinfo {pages} {163} (\bibinfo {year} {2009})}\BibitemShut {NoStop}%
\bibitem [{\citenamefont {Wright}(2010)}]{Wright2010}%
  \BibitemOpen
  \bibfield  {author} {\bibinfo {author} {\bibfnamefont {A.}~\bibnamefont
  {Wright}},\ }\href {\doibase 10.1038/nmat2637} {\bibfield  {journal}
  {\bibinfo  {journal} {Nat. Mater.}\ }\textbf {\bibinfo {volume} {9}},\
  \bibinfo {pages} {S5} (\bibinfo {year} {2010})}\BibitemShut {NoStop}%
\bibitem [{\citenamefont {Hentschel}\ \emph {et~al.}(2001)\citenamefont
  {Hentschel}, \citenamefont {Kienberger}, \citenamefont {Spielmann},
  \citenamefont {Reider}, \citenamefont {Milosevic}, \citenamefont {Brabec},
  \citenamefont {Corkum}, \citenamefont {Heinzmann}, \citenamefont {Drescher},\
  and\ \citenamefont {Krausz}}]{Hentschel2001}%
  \BibitemOpen
  \bibfield  {author} {\bibinfo {author} {\bibfnamefont {M.}~\bibnamefont
  {Hentschel}}, \bibinfo {author} {\bibfnamefont {R.}~\bibnamefont
  {Kienberger}}, \bibinfo {author} {\bibfnamefont {C.}~\bibnamefont
  {Spielmann}}, \bibinfo {author} {\bibfnamefont {G.~a.}\ \bibnamefont
  {Reider}}, \bibinfo {author} {\bibfnamefont {N.}~\bibnamefont {Milosevic}},
  \bibinfo {author} {\bibfnamefont {T.}~\bibnamefont {Brabec}}, \bibinfo
  {author} {\bibfnamefont {P.}~\bibnamefont {Corkum}}, \bibinfo {author}
  {\bibfnamefont {U.}~\bibnamefont {Heinzmann}}, \bibinfo {author}
  {\bibfnamefont {M.}~\bibnamefont {Drescher}}, \ and\ \bibinfo {author}
  {\bibfnamefont {F.}~\bibnamefont {Krausz}},\ }\href {\doibase
  10.1038/35107000} {\bibfield  {journal} {\bibinfo  {journal} {Nature}\
  }\textbf {\bibinfo {volume} {414}},\ \bibinfo {pages} {509} (\bibinfo {year}
  {2001})}\BibitemShut {NoStop}%
\bibitem [{\citenamefont {Smirnova}\ \emph {et~al.}(2009)\citenamefont
  {Smirnova}, \citenamefont {Mairesse}, \citenamefont {Patchkovskii},
  \citenamefont {Dudovich}, \citenamefont {Villeneuve}, \citenamefont
  {Corkum},\ and\ \citenamefont {Ivanov}}]{Smirnova2009}%
  \BibitemOpen
  \bibfield  {author} {\bibinfo {author} {\bibfnamefont {O.}~\bibnamefont
  {Smirnova}}, \bibinfo {author} {\bibfnamefont {Y.}~\bibnamefont {Mairesse}},
  \bibinfo {author} {\bibfnamefont {S.}~\bibnamefont {Patchkovskii}}, \bibinfo
  {author} {\bibfnamefont {N.}~\bibnamefont {Dudovich}}, \bibinfo {author}
  {\bibfnamefont {D.}~\bibnamefont {Villeneuve}}, \bibinfo {author}
  {\bibfnamefont {P.}~\bibnamefont {Corkum}}, \ and\ \bibinfo {author}
  {\bibfnamefont {M.~Y.}\ \bibnamefont {Ivanov}},\ }\href {\doibase
  10.1038/nature08253} {\bibfield  {journal} {\bibinfo  {journal} {Nature}\
  }\textbf {\bibinfo {volume} {460}},\ \bibinfo {pages} {972} (\bibinfo {year}
  {2009})}\BibitemShut {NoStop}%
\bibitem [{\citenamefont {W{\"{o}}rner}\ \emph {et~al.}(2010)\citenamefont
  {W{\"{o}}rner}, \citenamefont {Bertrand}, \citenamefont {Kartashov},
  \citenamefont {Corkum},\ and\ \citenamefont {Villeneuve}}]{Worner2010}%
  \BibitemOpen
  \bibfield  {author} {\bibinfo {author} {\bibfnamefont {H.~J.}\ \bibnamefont
  {W{\"{o}}rner}}, \bibinfo {author} {\bibfnamefont {J.~B.}\ \bibnamefont
  {Bertrand}}, \bibinfo {author} {\bibfnamefont {D.~V.}\ \bibnamefont
  {Kartashov}}, \bibinfo {author} {\bibfnamefont {P.~B.}\ \bibnamefont
  {Corkum}}, \ and\ \bibinfo {author} {\bibfnamefont {D.~M.}\ \bibnamefont
  {Villeneuve}},\ }\href {\doibase 10.1038/nature09185} {\bibfield  {journal}
  {\bibinfo  {journal} {Nature}\ }\textbf {\bibinfo {volume} {466}},\ \bibinfo
  {pages} {604} (\bibinfo {year} {2010})}\BibitemShut {NoStop}%
\bibitem [{\citenamefont {Morishita}\ \emph {et~al.}(2008)\citenamefont
  {Morishita}, \citenamefont {Le}, \citenamefont {Chen},\ and\ \citenamefont
  {Lin}}]{Morishita2008}%
  \BibitemOpen
  \bibfield  {author} {\bibinfo {author} {\bibfnamefont {T.}~\bibnamefont
  {Morishita}}, \bibinfo {author} {\bibfnamefont {A.-T.}\ \bibnamefont {Le}},
  \bibinfo {author} {\bibfnamefont {Z.}~\bibnamefont {Chen}}, \ and\ \bibinfo
  {author} {\bibfnamefont {C.~D.}\ \bibnamefont {Lin}},\ }\href {\doibase
  10.1103/PhysRevLett.100.013903} {\bibfield  {journal} {\bibinfo  {journal}
  {Phys. Rev. Lett.}\ }\textbf {\bibinfo {volume} {100}},\ \bibinfo {pages}
  {013903} (\bibinfo {year} {2008})},\ \Eprint {http://arxiv.org/abs/0707.3157}
  {arXiv:0707.3157} \BibitemShut {NoStop}%
\bibitem [{\citenamefont {Ghimire}\ \emph
  {et~al.}(2011{\natexlab{a}})\citenamefont {Ghimire}, \citenamefont
  {DiChiara}, \citenamefont {Sistrunk}, \citenamefont {Agostini}, \citenamefont
  {DiMauro},\ and\ \citenamefont {Reis}}]{Ghimire2010}%
  \BibitemOpen
  \bibfield  {author} {\bibinfo {author} {\bibfnamefont {S.}~\bibnamefont
  {Ghimire}}, \bibinfo {author} {\bibfnamefont {A.~D.}\ \bibnamefont
  {DiChiara}}, \bibinfo {author} {\bibfnamefont {E.}~\bibnamefont {Sistrunk}},
  \bibinfo {author} {\bibfnamefont {P.}~\bibnamefont {Agostini}}, \bibinfo
  {author} {\bibfnamefont {L.~F.}\ \bibnamefont {DiMauro}}, \ and\ \bibinfo
  {author} {\bibfnamefont {D.~A.}\ \bibnamefont {Reis}},\ }\href {\doibase
  10.1038/nphys1847} {\bibfield  {journal} {\bibinfo  {journal} {Nat. Phys.}\
  }\textbf {\bibinfo {volume} {7}},\ \bibinfo {pages} {138} (\bibinfo {year}
  {2011}{\natexlab{a}})}\BibitemShut {NoStop}%
\bibitem [{\citenamefont {Ghimire}\ \emph
  {et~al.}(2011{\natexlab{b}})\citenamefont {Ghimire}, \citenamefont
  {DiChiara}, \citenamefont {Sistrunk}, \citenamefont {Szafruga}, \citenamefont
  {Agostini}, \citenamefont {DiMauro},\ and\ \citenamefont
  {Reis}}]{Ghimire2011}%
  \BibitemOpen
  \bibfield  {author} {\bibinfo {author} {\bibfnamefont {S.}~\bibnamefont
  {Ghimire}}, \bibinfo {author} {\bibfnamefont {A.~D.}\ \bibnamefont
  {DiChiara}}, \bibinfo {author} {\bibfnamefont {E.}~\bibnamefont {Sistrunk}},
  \bibinfo {author} {\bibfnamefont {U.~B.}\ \bibnamefont {Szafruga}}, \bibinfo
  {author} {\bibfnamefont {P.}~\bibnamefont {Agostini}}, \bibinfo {author}
  {\bibfnamefont {L.~F.}\ \bibnamefont {DiMauro}}, \ and\ \bibinfo {author}
  {\bibfnamefont {D.~A.}\ \bibnamefont {Reis}},\ }\href {\doibase
  10.1103/PhysRevLett.107.167407} {\bibfield  {journal} {\bibinfo  {journal}
  {Phys. Rev. Lett.}\ }\textbf {\bibinfo {volume} {107}},\ \bibinfo {pages}
  {167407} (\bibinfo {year} {2011}{\natexlab{b}})}\BibitemShut {NoStop}%
\bibitem [{\citenamefont {Schubert}\ \emph {et~al.}(2014)\citenamefont
  {Schubert}, \citenamefont {Hohenleutner}, \citenamefont {Langer},
  \citenamefont {Urbanek}, \citenamefont {Lange}, \citenamefont {Huttner},
  \citenamefont {Golde}, \citenamefont {Meier}, \citenamefont {Kira},
  \citenamefont {Koch},\ and\ \citenamefont {Huber}}]{Schubert2014}%
  \BibitemOpen
  \bibfield  {author} {\bibinfo {author} {\bibfnamefont {O.}~\bibnamefont
  {Schubert}}, \bibinfo {author} {\bibfnamefont {M.}~\bibnamefont
  {Hohenleutner}}, \bibinfo {author} {\bibfnamefont {F.}~\bibnamefont
  {Langer}}, \bibinfo {author} {\bibfnamefont {B.}~\bibnamefont {Urbanek}},
  \bibinfo {author} {\bibfnamefont {C.}~\bibnamefont {Lange}}, \bibinfo
  {author} {\bibfnamefont {U.}~\bibnamefont {Huttner}}, \bibinfo {author}
  {\bibfnamefont {D.}~\bibnamefont {Golde}}, \bibinfo {author} {\bibfnamefont
  {T.}~\bibnamefont {Meier}}, \bibinfo {author} {\bibfnamefont
  {M.}~\bibnamefont {Kira}}, \bibinfo {author} {\bibfnamefont {S.~W.}\
  \bibnamefont {Koch}}, \ and\ \bibinfo {author} {\bibfnamefont
  {R.}~\bibnamefont {Huber}},\ }\href {\doibase 10.1038/nphoton.2013.349}
  {\bibfield  {journal} {\bibinfo  {journal} {Nat. Photonics}\ }\textbf
  {\bibinfo {volume} {8}},\ \bibinfo {pages} {119} (\bibinfo {year}
  {2014})}\BibitemShut {NoStop}%
\bibitem [{\citenamefont {Hohenleutner}\ \emph {et~al.}(2015)\citenamefont
  {Hohenleutner}, \citenamefont {Langer}, \citenamefont {Schubert},
  \citenamefont {Knorr}, \citenamefont {Huttner}, \citenamefont {Koch},
  \citenamefont {Kira},\ and\ \citenamefont {Huber}}]{Hohenleutner2015}%
  \BibitemOpen
  \bibfield  {author} {\bibinfo {author} {\bibfnamefont {M.}~\bibnamefont
  {Hohenleutner}}, \bibinfo {author} {\bibfnamefont {F.}~\bibnamefont
  {Langer}}, \bibinfo {author} {\bibfnamefont {O.}~\bibnamefont {Schubert}},
  \bibinfo {author} {\bibfnamefont {M.}~\bibnamefont {Knorr}}, \bibinfo
  {author} {\bibfnamefont {U.}~\bibnamefont {Huttner}}, \bibinfo {author}
  {\bibfnamefont {S.~W.}\ \bibnamefont {Koch}}, \bibinfo {author}
  {\bibfnamefont {M.}~\bibnamefont {Kira}}, \ and\ \bibinfo {author}
  {\bibfnamefont {R.}~\bibnamefont {Huber}},\ }\href {\doibase
  10.1038/nature14652} {\bibfield  {journal} {\bibinfo  {journal} {Nature}\
  }\textbf {\bibinfo {volume} {523}},\ \bibinfo {pages} {572} (\bibinfo {year}
  {2015})}\BibitemShut {NoStop}%
\bibitem [{\citenamefont {Mahmood}\ \emph {et~al.}(2016)\citenamefont
  {Mahmood}, \citenamefont {Chan}, \citenamefont {Alpichshev}, \citenamefont
  {Gardner}, \citenamefont {Lee}, \citenamefont {Lee},\ and\ \citenamefont
  {Gedik}}]{Mahmood2015}%
  \BibitemOpen
  \bibfield  {author} {\bibinfo {author} {\bibfnamefont {F.}~\bibnamefont
  {Mahmood}}, \bibinfo {author} {\bibfnamefont {C.-K.}\ \bibnamefont {Chan}},
  \bibinfo {author} {\bibfnamefont {Z.}~\bibnamefont {Alpichshev}}, \bibinfo
  {author} {\bibfnamefont {D.}~\bibnamefont {Gardner}}, \bibinfo {author}
  {\bibfnamefont {Y.}~\bibnamefont {Lee}}, \bibinfo {author} {\bibfnamefont
  {P.~A.}\ \bibnamefont {Lee}}, \ and\ \bibinfo {author} {\bibfnamefont
  {N.}~\bibnamefont {Gedik}},\ }\href {\doibase 10.1038/nphys3609} {\bibfield
  {journal} {\bibinfo  {journal} {Nat. Phys.}\ }\textbf {\bibinfo {volume}
  {12}},\ \bibinfo {pages} {306} (\bibinfo {year} {2016})},\ \Eprint
  {http://arxiv.org/abs/1512.05714} {arXiv:1512.05714} \BibitemShut {NoStop}%
\bibitem [{\citenamefont {Vampa}\ \emph
  {et~al.}(2015{\natexlab{a}})\citenamefont {Vampa}, \citenamefont {Hammond},
  \citenamefont {Thir{\'{e}}}, \citenamefont {Schmidt}, \citenamefont
  {L{\'{e}}gar{\'{e}}}, \citenamefont {McDonald}, \citenamefont {Brabec},\ and\
  \citenamefont {Corkum}}]{Vampa2015a}%
  \BibitemOpen
  \bibfield  {author} {\bibinfo {author} {\bibfnamefont {G.}~\bibnamefont
  {Vampa}}, \bibinfo {author} {\bibfnamefont {T.~J.}\ \bibnamefont {Hammond}},
  \bibinfo {author} {\bibfnamefont {N.}~\bibnamefont {Thir{\'{e}}}}, \bibinfo
  {author} {\bibfnamefont {B.~E.}\ \bibnamefont {Schmidt}}, \bibinfo {author}
  {\bibfnamefont {F.}~\bibnamefont {L{\'{e}}gar{\'{e}}}}, \bibinfo {author}
  {\bibfnamefont {C.~R.}\ \bibnamefont {McDonald}}, \bibinfo {author}
  {\bibfnamefont {T.}~\bibnamefont {Brabec}}, \ and\ \bibinfo {author}
  {\bibfnamefont {P.~B.}\ \bibnamefont {Corkum}},\ }\href {\doibase
  10.1038/nature14517} {\bibfield  {journal} {\bibinfo  {journal} {Nature}\
  }\textbf {\bibinfo {volume} {522}},\ \bibinfo {pages} {462} (\bibinfo {year}
  {2015}{\natexlab{a}})}\BibitemShut {NoStop}%
\bibitem [{\citenamefont {Ndabashimiye}\ \emph {et~al.}(2016)\citenamefont
  {Ndabashimiye}, \citenamefont {Ghimire}, \citenamefont {Wu}, \citenamefont
  {Browne}, \citenamefont {Schafer}, \citenamefont {Gaarde},\ and\
  \citenamefont {Reis}}]{NdabashimiyeG.2016a}%
  \BibitemOpen
  \bibfield  {author} {\bibinfo {author} {\bibfnamefont {G.}~\bibnamefont
  {Ndabashimiye}}, \bibinfo {author} {\bibfnamefont {S.}~\bibnamefont
  {Ghimire}}, \bibinfo {author} {\bibfnamefont {M.}~\bibnamefont {Wu}},
  \bibinfo {author} {\bibfnamefont {D.~A.}\ \bibnamefont {Browne}}, \bibinfo
  {author} {\bibfnamefont {K.~J.}\ \bibnamefont {Schafer}}, \bibinfo {author}
  {\bibfnamefont {M.~B.}\ \bibnamefont {Gaarde}}, \ and\ \bibinfo {author}
  {\bibfnamefont {D.~A.}\ \bibnamefont {Reis}},\ }\href {\doibase
  10.1038/nature17660} {\bibfield  {journal} {\bibinfo  {journal} {Nature}\
  }\textbf {\bibinfo {volume} {534}},\ \bibinfo {pages} {520} (\bibinfo {year}
  {2016})}\BibitemShut {NoStop}%
\bibitem [{\citenamefont {Korbman}\ \emph {et~al.}(2013)\citenamefont
  {Korbman}, \citenamefont {{Yu Kruchinin}},\ and\ \citenamefont
  {Yakovlev}}]{Korbman2013}%
  \BibitemOpen
  \bibfield  {author} {\bibinfo {author} {\bibfnamefont {M.}~\bibnamefont
  {Korbman}}, \bibinfo {author} {\bibfnamefont {S.}~\bibnamefont {{Yu
  Kruchinin}}}, \ and\ \bibinfo {author} {\bibfnamefont {V.~S.}\ \bibnamefont
  {Yakovlev}},\ }\href {\doibase 10.1088/1367-2630/15/1/013006} {\bibfield
  {journal} {\bibinfo  {journal} {New J. Phys.}\ }\textbf {\bibinfo {volume}
  {15}},\ \bibinfo {pages} {013006} (\bibinfo {year} {2013})}\BibitemShut
  {NoStop}%
\bibitem [{\citenamefont {Kemper}\ \emph {et~al.}(2013)\citenamefont {Kemper},
  \citenamefont {Moritz}, \citenamefont {Freericks},\ and\ \citenamefont
  {Devereaux}}]{Kemper2013}%
  \BibitemOpen
  \bibfield  {author} {\bibinfo {author} {\bibfnamefont {A.~F.}\ \bibnamefont
  {Kemper}}, \bibinfo {author} {\bibfnamefont {B.}~\bibnamefont {Moritz}},
  \bibinfo {author} {\bibfnamefont {J.~K.}\ \bibnamefont {Freericks}}, \ and\
  \bibinfo {author} {\bibfnamefont {T.~P.}\ \bibnamefont {Devereaux}},\ }\href
  {\doibase 10.1088/1367-2630/15/2/023003} {\bibfield  {journal} {\bibinfo
  {journal} {New J. Phys.}\ }\textbf {\bibinfo {volume} {15}},\ \bibinfo
  {pages} {023003} (\bibinfo {year} {2013})}\BibitemShut {NoStop}%
\bibitem [{\citenamefont {Vampa}\ \emph {et~al.}(2014)\citenamefont {Vampa},
  \citenamefont {McDonald}, \citenamefont {Orlando}, \citenamefont {Klug},
  \citenamefont {Corkum},\ and\ \citenamefont {Brabec}}]{Vampa2014}%
  \BibitemOpen
  \bibfield  {author} {\bibinfo {author} {\bibfnamefont {G.}~\bibnamefont
  {Vampa}}, \bibinfo {author} {\bibfnamefont {C.~R.}\ \bibnamefont {McDonald}},
  \bibinfo {author} {\bibfnamefont {G.}~\bibnamefont {Orlando}}, \bibinfo
  {author} {\bibfnamefont {D.~D.}\ \bibnamefont {Klug}}, \bibinfo {author}
  {\bibfnamefont {P.~B.}\ \bibnamefont {Corkum}}, \ and\ \bibinfo {author}
  {\bibfnamefont {T.}~\bibnamefont {Brabec}},\ }\href {\doibase
  10.1103/PhysRevLett.113.073901} {\bibfield  {journal} {\bibinfo  {journal}
  {Phys. Rev. Lett.}\ }\textbf {\bibinfo {volume} {113}},\ \bibinfo {pages}
  {073901} (\bibinfo {year} {2014})}\BibitemShut {NoStop}%
\bibitem [{\citenamefont {Higuchi}\ \emph {et~al.}(2014)\citenamefont
  {Higuchi}, \citenamefont {Stockman},\ and\ \citenamefont
  {Hommelhoff}}]{Higuchi2014a}%
  \BibitemOpen
  \bibfield  {author} {\bibinfo {author} {\bibfnamefont {T.}~\bibnamefont
  {Higuchi}}, \bibinfo {author} {\bibfnamefont {M.~I.}\ \bibnamefont
  {Stockman}}, \ and\ \bibinfo {author} {\bibfnamefont {P.}~\bibnamefont
  {Hommelhoff}},\ }\href {\doibase 10.1103/PhysRevLett.113.213901} {\bibfield
  {journal} {\bibinfo  {journal} {Phys. Rev. Lett.}\ }\textbf {\bibinfo
  {volume} {113}},\ \bibinfo {pages} {213901} (\bibinfo {year}
  {2014})}\BibitemShut {NoStop}%
\bibitem [{\citenamefont {Hawkins}\ \emph {et~al.}(2015)\citenamefont
  {Hawkins}, \citenamefont {Ivanov},\ and\ \citenamefont
  {Yakovlev}}]{Hawkins2015}%
  \BibitemOpen
  \bibfield  {author} {\bibinfo {author} {\bibfnamefont {P.~G.}\ \bibnamefont
  {Hawkins}}, \bibinfo {author} {\bibfnamefont {M.~Y.}\ \bibnamefont {Ivanov}},
  \ and\ \bibinfo {author} {\bibfnamefont {V.~S.}\ \bibnamefont {Yakovlev}},\
  }\href {\doibase 10.1103/PhysRevA.91.013405} {\bibfield  {journal} {\bibinfo
  {journal} {Phys. Rev. A}\ }\textbf {\bibinfo {volume} {91}},\ \bibinfo
  {pages} {013405} (\bibinfo {year} {2015})}\BibitemShut {NoStop}%
\bibitem [{\citenamefont {Wu}\ \emph {et~al.}(2015)\citenamefont {Wu},
  \citenamefont {Ghimire}, \citenamefont {Reis}, \citenamefont {Schafer},\ and\
  \citenamefont {Gaarde}}]{Wu2015}%
  \BibitemOpen
  \bibfield  {author} {\bibinfo {author} {\bibfnamefont {M.}~\bibnamefont
  {Wu}}, \bibinfo {author} {\bibfnamefont {S.}~\bibnamefont {Ghimire}},
  \bibinfo {author} {\bibfnamefont {D.~A.}\ \bibnamefont {Reis}}, \bibinfo
  {author} {\bibfnamefont {K.~J.}\ \bibnamefont {Schafer}}, \ and\ \bibinfo
  {author} {\bibfnamefont {M.~B.}\ \bibnamefont {Gaarde}},\ }\href {\doibase
  10.1103/PhysRevA.91.043839} {\bibfield  {journal} {\bibinfo  {journal} {Phys.
  Rev. A}\ }\textbf {\bibinfo {volume} {91}},\ \bibinfo {pages} {043839}
  (\bibinfo {year} {2015})}\BibitemShut {NoStop}%
\bibitem [{\citenamefont {Vampa}\ \emph
  {et~al.}(2015{\natexlab{b}})\citenamefont {Vampa}, \citenamefont {McDonald},
  \citenamefont {Fraser},\ and\ \citenamefont {Brabec}}]{Vampa2015b}%
  \BibitemOpen
  \bibfield  {author} {\bibinfo {author} {\bibfnamefont {G.}~\bibnamefont
  {Vampa}}, \bibinfo {author} {\bibfnamefont {C.}~\bibnamefont {McDonald}},
  \bibinfo {author} {\bibfnamefont {A.}~\bibnamefont {Fraser}}, \ and\ \bibinfo
  {author} {\bibfnamefont {T.}~\bibnamefont {Brabec}},\ }\href {\doibase
  10.1109/JSTQE.2015.2402636} {\bibfield  {journal} {\bibinfo  {journal} {IEEE
  J. Sel. Top. Quantum Electron.}\ }\textbf {\bibinfo {volume} {21}} (\bibinfo
  {year} {2015}{\natexlab{b}}),\ 10.1109/JSTQE.2015.2402636}\BibitemShut
  {NoStop}%
\bibitem [{\citenamefont {Vampa}\ \emph
  {et~al.}(2015{\natexlab{c}})\citenamefont {Vampa}, \citenamefont {McDonald},
  \citenamefont {Orlando}, \citenamefont {Corkum},\ and\ \citenamefont
  {Brabec}}]{Vampa2015}%
  \BibitemOpen
  \bibfield  {author} {\bibinfo {author} {\bibfnamefont {G.}~\bibnamefont
  {Vampa}}, \bibinfo {author} {\bibfnamefont {C.~R.}\ \bibnamefont {McDonald}},
  \bibinfo {author} {\bibfnamefont {G.}~\bibnamefont {Orlando}}, \bibinfo
  {author} {\bibfnamefont {P.~B.}\ \bibnamefont {Corkum}}, \ and\ \bibinfo
  {author} {\bibfnamefont {T.}~\bibnamefont {Brabec}},\ }\href {\doibase
  10.1103/PhysRevB.91.064302} {\bibfield  {journal} {\bibinfo  {journal} {Phys.
  Rev. B}\ }\textbf {\bibinfo {volume} {91}},\ \bibinfo {pages} {064302}
  (\bibinfo {year} {2015}{\natexlab{c}})}\BibitemShut {NoStop}%
\bibitem [{\citenamefont {McDonald}\ \emph {et~al.}(2015)\citenamefont
  {McDonald}, \citenamefont {Vampa}, \citenamefont {Corkum},\ and\
  \citenamefont {Brabec}}]{McDonald2015}%
  \BibitemOpen
  \bibfield  {author} {\bibinfo {author} {\bibfnamefont {C.~R.}\ \bibnamefont
  {McDonald}}, \bibinfo {author} {\bibfnamefont {G.}~\bibnamefont {Vampa}},
  \bibinfo {author} {\bibfnamefont {P.~B.}\ \bibnamefont {Corkum}}, \ and\
  \bibinfo {author} {\bibfnamefont {T.}~\bibnamefont {Brabec}},\ }\href
  {\doibase 10.1103/PhysRevA.92.033845} {\bibfield  {journal} {\bibinfo
  {journal} {Phys. Rev. A}\ }\textbf {\bibinfo {volume} {92}},\ \bibinfo
  {pages} {033845} (\bibinfo {year} {2015})}\BibitemShut {NoStop}%
\bibitem [{\citenamefont {Guan}\ \emph {et~al.}(2016)\citenamefont {Guan},
  \citenamefont {Zhou},\ and\ \citenamefont {Bian}}]{Guan2015}%
  \BibitemOpen
  \bibfield  {author} {\bibinfo {author} {\bibfnamefont {Z.}~\bibnamefont
  {Guan}}, \bibinfo {author} {\bibfnamefont {X.-X.}\ \bibnamefont {Zhou}}, \
  and\ \bibinfo {author} {\bibfnamefont {X.-B.}\ \bibnamefont {Bian}},\ }\href
  {\doibase 10.1103/PhysRevA.93.033852} {\bibfield  {journal} {\bibinfo
  {journal} {Phys. Rev. A}\ }\textbf {\bibinfo {volume} {93}},\ \bibinfo
  {pages} {033852} (\bibinfo {year} {2016})},\ \Eprint
  {http://arxiv.org/abs/1511.08398} {arXiv:1511.08398} \BibitemShut {NoStop}%
\bibitem [{\citenamefont {Vampa}\ \emph
  {et~al.}(2015{\natexlab{d}})\citenamefont {Vampa}, \citenamefont {Hammond},
  \citenamefont {Thir{\'{e}}}, \citenamefont {Schmidt}, \citenamefont
  {L{\'{e}}gar{\'{e}}}, \citenamefont {McDonald}, \citenamefont {Brabec},
  \citenamefont {Klug},\ and\ \citenamefont {Corkum}}]{Vampa2015c}%
  \BibitemOpen
  \bibfield  {author} {\bibinfo {author} {\bibfnamefont {G.}~\bibnamefont
  {Vampa}}, \bibinfo {author} {\bibfnamefont {T.~J.}\ \bibnamefont {Hammond}},
  \bibinfo {author} {\bibfnamefont {N.}~\bibnamefont {Thir{\'{e}}}}, \bibinfo
  {author} {\bibfnamefont {B.~E.}\ \bibnamefont {Schmidt}}, \bibinfo {author}
  {\bibfnamefont {F.}~\bibnamefont {L{\'{e}}gar{\'{e}}}}, \bibinfo {author}
  {\bibfnamefont {C.~R.}\ \bibnamefont {McDonald}}, \bibinfo {author}
  {\bibfnamefont {T.}~\bibnamefont {Brabec}}, \bibinfo {author} {\bibfnamefont
  {D.~D.}\ \bibnamefont {Klug}}, \ and\ \bibinfo {author} {\bibfnamefont
  {P.~B.}\ \bibnamefont {Corkum}},\ }\href {\doibase
  10.1103/PhysRevLett.115.193603} {\bibfield  {journal} {\bibinfo  {journal}
  {Phys. Rev. Lett.}\ }\textbf {\bibinfo {volume} {115}},\ \bibinfo {pages}
  {193603} (\bibinfo {year} {2015}{\natexlab{d}})}\BibitemShut {NoStop}%
\bibitem [{\citenamefont {Krieger}\ and\ \citenamefont
  {Iafrate}(1986)}]{Krieger1986}%
  \BibitemOpen
  \bibfield  {author} {\bibinfo {author} {\bibfnamefont {J.~B.}\ \bibnamefont
  {Krieger}}\ and\ \bibinfo {author} {\bibfnamefont {G.~J.}\ \bibnamefont
  {Iafrate}},\ }\href {\doibase 10.1103/PhysRevB.33.5494} {\bibfield  {journal}
  {\bibinfo  {journal} {Phys. Rev. B}\ }\textbf {\bibinfo {volume} {33}},\
  \bibinfo {pages} {5494} (\bibinfo {year} {1986})}\BibitemShut {NoStop}%
\bibitem [{\citenamefont {Krainov}\ and\ \citenamefont
  {Yakovlev}(1980)}]{KraTnov1980}%
  \BibitemOpen
  \bibfield  {author} {\bibinfo {author} {\bibfnamefont {V.~P.}\ \bibnamefont
  {Krainov}}\ and\ \bibinfo {author} {\bibfnamefont {V.~P.}\ \bibnamefont
  {Yakovlev}},\ }\href
  {http://www.jetp.ac.ru/cgi-bin/dn/e{\_}051{\_}06{\_}1104.pdf} {\bibfield
  {journal} {\bibinfo  {journal} {Zh. Eksp. Teor. Fiz}\ }\textbf {\bibinfo
  {volume} {51}},\ \bibinfo {pages} {1104} (\bibinfo {year}
  {1980})}\BibitemShut {NoStop}%
\bibitem [{\citenamefont {Averbukh}\ and\ \citenamefont
  {Perelman}(1985)}]{Averbukh1985}%
  \BibitemOpen
  \bibfield  {author} {\bibinfo {author} {\bibfnamefont {I.~S.}\ \bibnamefont
  {Averbukh}}\ and\ \bibinfo {author} {\bibfnamefont {N.~F.}\ \bibnamefont
  {Perelman}},\ }\href
  {http://www.jetp.ac.ru/cgi-bin/dn/e{\_}061{\_}04{\_}0665.pdf} {\bibfield
  {journal} {\bibinfo  {journal} {Sov. J. Exp. Theor. Phys.}\ }\textbf
  {\bibinfo {volume} {61}},\ \bibinfo {pages} {665} (\bibinfo {year}
  {1985})}\BibitemShut {NoStop}%
\bibitem [{\citenamefont {Plaja}\ and\ \citenamefont
  {Roso-Franco}(1992)}]{Plaja1992}%
  \BibitemOpen
  \bibfield  {author} {\bibinfo {author} {\bibfnamefont {L.}~\bibnamefont
  {Plaja}}\ and\ \bibinfo {author} {\bibfnamefont {L.}~\bibnamefont
  {Roso-Franco}},\ }\href {\doibase 10.1364/JOSAB.9.002210} {\bibfield
  {journal} {\bibinfo  {journal} {J. Opt. Soc. Am. B}\ }\textbf {\bibinfo
  {volume} {9}},\ \bibinfo {pages} {2210} (\bibinfo {year} {1992})}\BibitemShut
  {NoStop}%
\bibitem [{\citenamefont {Ivanov}\ \emph {et~al.}(1993)\citenamefont {Ivanov},
  \citenamefont {Corkum},\ and\ \citenamefont {Dietrich}}]{Ivanov1993}%
  \BibitemOpen
  \bibfield  {author} {\bibinfo {author} {\bibfnamefont {M.~Y.}\ \bibnamefont
  {Ivanov}}, \bibinfo {author} {\bibfnamefont {P.~B.}\ \bibnamefont {Corkum}},
  \ and\ \bibinfo {author} {\bibfnamefont {P.}~\bibnamefont {Dietrich}},\
  }\href@noop {} {\bibfield  {journal} {\bibinfo  {journal} {Laser Phys.}\
  }\textbf {\bibinfo {volume} {3}},\ \bibinfo {pages} {375} (\bibinfo {year}
  {1993})}\BibitemShut {NoStop}%
\bibitem [{\citenamefont {Krainov}\ and\ \citenamefont
  {Mulyukov}(1994)}]{Krainov1994}%
  \BibitemOpen
  \bibfield  {author} {\bibinfo {author} {\bibfnamefont {V.}~\bibnamefont
  {Krainov}}\ and\ \bibinfo {author} {\bibfnamefont {Z.}~\bibnamefont
  {Mulyukov}},\ }\href
  {http://www.maik.ru/full/lasphys{\_}archive/94/3/lasphys3{\_}94p544full.pdf}
  {\bibfield  {journal} {\bibinfo  {journal} {Laser Phys}\ }\textbf {\bibinfo
  {volume} {4}},\ \bibinfo {pages} {544} (\bibinfo {year} {1994})}\BibitemShut
  {NoStop}%
\bibitem [{\citenamefont {Gauthey}\ \emph {et~al.}(1997)\citenamefont
  {Gauthey}, \citenamefont {Garraway},\ and\ \citenamefont
  {Knight}}]{Gauthey1997a}%
  \BibitemOpen
  \bibfield  {author} {\bibinfo {author} {\bibfnamefont {F.~I.}\ \bibnamefont
  {Gauthey}}, \bibinfo {author} {\bibfnamefont {B.~M.}\ \bibnamefont
  {Garraway}}, \ and\ \bibinfo {author} {\bibfnamefont {P.~L.}\ \bibnamefont
  {Knight}},\ }\href {\doibase 10.1103/PhysRevA.56.3093} {\bibfield  {journal}
  {\bibinfo  {journal} {Phys. Rev. A}\ }\textbf {\bibinfo {volume} {56}},\
  \bibinfo {pages} {3093} (\bibinfo {year} {1997})}\BibitemShut {NoStop}%
\bibitem [{\citenamefont {Santana}\ \emph {et~al.}(2001)\citenamefont
  {Santana}, \citenamefont {Llorente},\ and\ \citenamefont
  {Delgado}}]{Santana2000}%
  \BibitemOpen
  \bibfield  {author} {\bibinfo {author} {\bibfnamefont {A.}~\bibnamefont
  {Santana}}, \bibinfo {author} {\bibfnamefont {J.~M.~G.}\ \bibnamefont
  {Llorente}}, \ and\ \bibinfo {author} {\bibfnamefont {V.}~\bibnamefont
  {Delgado}},\ }\href {\doibase 10.1088/0953-4075/34/12/306} {\bibfield
  {journal} {\bibinfo  {journal} {J. Phys. B At. Mol. Opt. Phys.}\ }\textbf
  {\bibinfo {volume} {34}},\ \bibinfo {pages} {2371} (\bibinfo {year}
  {2001})}\BibitemShut {NoStop}%
\bibitem [{\citenamefont {{Figueira de Morisson Faria}}\ and\ \citenamefont
  {Rotter}(2002)}]{FigueiradeMorissonFaria2002}%
  \BibitemOpen
  \bibfield  {author} {\bibinfo {author} {\bibfnamefont {C.}~\bibnamefont
  {{Figueira de Morisson Faria}}}\ and\ \bibinfo {author} {\bibfnamefont
  {I.}~\bibnamefont {Rotter}},\ }\href {\doibase 10.1103/PhysRevA.66.013402}
  {\bibfield  {journal} {\bibinfo  {journal} {Phys. Rev. A}\ }\textbf {\bibinfo
  {volume} {66}},\ \bibinfo {pages} {013402} (\bibinfo {year}
  {2002})}\BibitemShut {NoStop}%
\bibitem [{\citenamefont {Haljan}\ \emph {et~al.}(2003)\citenamefont {Haljan},
  \citenamefont {Fortier}, \citenamefont {Hawrylak}, \citenamefont {Corkum},
  \citenamefont {Ivanov}, \citenamefont {Nrc}, \citenamefont {Rd},\ and\
  \citenamefont {Or}}]{Haljan2003}%
  \BibitemOpen
  \bibfield  {author} {\bibinfo {author} {\bibfnamefont {P.}~\bibnamefont
  {Haljan}}, \bibinfo {author} {\bibfnamefont {T.}~\bibnamefont {Fortier}},
  \bibinfo {author} {\bibfnamefont {P.}~\bibnamefont {Hawrylak}}, \bibinfo
  {author} {\bibfnamefont {P.~B.}\ \bibnamefont {Corkum}}, \bibinfo {author}
  {\bibfnamefont {M.~Y.}\ \bibnamefont {Ivanov}}, \bibinfo {author}
  {\bibfnamefont {I.~M.~S.}\ \bibnamefont {Nrc}}, \bibinfo {author}
  {\bibfnamefont {M.-M.}\ \bibnamefont {Rd}}, \ and\ \bibinfo {author}
  {\bibfnamefont {O.~K.~a.}\ \bibnamefont {Or}},\ }\href@noop {} {\bibfield
  {journal} {\bibinfo  {journal} {Laser Phys.}\ }\textbf {\bibinfo {volume}
  {13}},\ \bibinfo {pages} {452} (\bibinfo {year} {2003})}\BibitemShut
  {NoStop}%
\bibitem [{\citenamefont {Faria}\ and\ \citenamefont
  {Rotter}(2003)}]{Faria2003}%
  \BibitemOpen
  \bibfield  {author} {\bibinfo {author} {\bibfnamefont {C.~d.~M.}\
  \bibnamefont {Faria}}\ and\ \bibinfo {author} {\bibfnamefont
  {I.}~\bibnamefont {Rotter}},\ }\href
  {http://www.homepages.ucl.ac.uk/{~}ucapcfi/lp.pdf} {\bibfield  {journal}
  {\bibinfo  {journal} {LASER Phys.}\ }\textbf {\bibinfo {volume} {13}},\
  \bibinfo {pages} {985} (\bibinfo {year} {2003})}\BibitemShut {NoStop}%
\bibitem [{\citenamefont {Ashhab}\ \emph {et~al.}(2007)\citenamefont {Ashhab},
  \citenamefont {Johansson}, \citenamefont {Zagoskin},\ and\ \citenamefont
  {Nori}}]{Ashhab2007}%
  \BibitemOpen
  \bibfield  {author} {\bibinfo {author} {\bibfnamefont {S.}~\bibnamefont
  {Ashhab}}, \bibinfo {author} {\bibfnamefont {J.~R.}\ \bibnamefont
  {Johansson}}, \bibinfo {author} {\bibfnamefont {A.~M.}\ \bibnamefont
  {Zagoskin}}, \ and\ \bibinfo {author} {\bibfnamefont {F.}~\bibnamefont
  {Nori}},\ }\href {\doibase 10.1103/PhysRevA.75.063414} {\bibfield  {journal}
  {\bibinfo  {journal} {Phys. Rev. A}\ }\textbf {\bibinfo {volume} {75}},\
  \bibinfo {pages} {063414} (\bibinfo {year} {2007})},\ \Eprint
  {http://arxiv.org/abs/0702032} {arXiv:0702032 [quant-ph]} \BibitemShut
  {NoStop}%
\bibitem [{\citenamefont {Pic{\'{o}}n}\ \emph {et~al.}(2010)\citenamefont
  {Pic{\'{o}}n}, \citenamefont {Roso}, \citenamefont {Mompart}, \citenamefont
  {Varela}, \citenamefont {Ahufinger}, \citenamefont {Corbal{\'{a}}n},\ and\
  \citenamefont {Plaja}}]{Pic??n2010}%
  \BibitemOpen
  \bibfield  {author} {\bibinfo {author} {\bibfnamefont {A.}~\bibnamefont
  {Pic{\'{o}}n}}, \bibinfo {author} {\bibfnamefont {L.}~\bibnamefont {Roso}},
  \bibinfo {author} {\bibfnamefont {J.}~\bibnamefont {Mompart}}, \bibinfo
  {author} {\bibfnamefont {O.}~\bibnamefont {Varela}}, \bibinfo {author}
  {\bibfnamefont {V.}~\bibnamefont {Ahufinger}}, \bibinfo {author}
  {\bibfnamefont {R.}~\bibnamefont {Corbal{\'{a}}n}}, \ and\ \bibinfo {author}
  {\bibfnamefont {L.}~\bibnamefont {Plaja}},\ }\href {\doibase
  10.1103/PhysRevA.81.033420} {\bibfield  {journal} {\bibinfo  {journal} {Phys.
  Rev. A}\ }\textbf {\bibinfo {volume} {81}},\ \bibinfo {pages} {033420}
  (\bibinfo {year} {2010})},\ \Eprint {http://arxiv.org/abs/1003.4566}
  {arXiv:1003.4566} \BibitemShut {NoStop}%
\bibitem [{\citenamefont {Mkrtchian}\ \emph {et~al.}(2012)\citenamefont
  {Mkrtchian}, \citenamefont {Avchyan},\ and\ \citenamefont
  {F}}]{Mkrtchian2012}%
  \BibitemOpen
  \bibfield  {author} {\bibinfo {author} {\bibfnamefont {H.~K.~A.}\
  \bibnamefont {Mkrtchian}}, \bibinfo {author} {\bibfnamefont {B.~R.}\
  \bibnamefont {Avchyan}}, \ and\ \bibinfo {author} {\bibfnamefont
  {G.}~\bibnamefont {F}},\ }\href {\doibase 10.1088/0953-4075/45/2/025402}
  {\bibfield  {journal} {\bibinfo  {journal} {J. Phys. B At. Mol. Opt. Phys.}\
  }\textbf {\bibinfo {volume} {45}},\ \bibinfo {pages} {25402} (\bibinfo {year}
  {2012})}\BibitemShut {NoStop}%
\bibitem [{\citenamefont {Garraway}\ and\ \citenamefont
  {Stenholm}(1992)}]{Garraway1992}%
  \BibitemOpen
  \bibfield  {author} {\bibinfo {author} {\bibfnamefont {B.~M.}\ \bibnamefont
  {Garraway}}\ and\ \bibinfo {author} {\bibfnamefont {S.}~\bibnamefont
  {Stenholm}},\ }\href {\doibase 10.1103/PhysRevA.45.364} {\bibfield  {journal}
  {\bibinfo  {journal} {Phys. Rev. A}\ }\textbf {\bibinfo {volume} {45}},\
  \bibinfo {pages} {364} (\bibinfo {year} {1992})}\BibitemShut {NoStop}%
\bibitem [{\citenamefont {{Wolfram Research Inc.}}(2016)}]{Mathematica}%
  \BibitemOpen
  \bibfield  {author} {\bibinfo {author} {\bibnamefont {{Wolfram Research
  Inc.}}},\ }\href {http://www.wolfram.com} {\emph {\bibinfo {title}
  {Mathematica 11.0}}} (\bibinfo {year} {2016}),\ \bibinfo {note} {see
  ContinuousWaveletTransform}\BibitemShut {NoStop}%
\bibitem [{\citenamefont {Slater}(1952)}]{Slater1952}%
  \BibitemOpen
  \bibfield  {author} {\bibinfo {author} {\bibfnamefont {J.~C.}\ \bibnamefont
  {Slater}},\ }\href {\doibase 10.1103/PhysRev.87.807} {\bibfield  {journal}
  {\bibinfo  {journal} {Phys. Rev.}\ }\textbf {\bibinfo {volume} {87}},\
  \bibinfo {pages} {807} (\bibinfo {year} {1952})}\BibitemShut {NoStop}%
\bibitem [{\citenamefont {Ashcroft}\ and\ \citenamefont
  {Mermin}(1976)}]{Ashcroft1976}%
  \BibitemOpen
  \bibfield  {author} {\bibinfo {author} {\bibfnamefont {N.~W.}\ \bibnamefont
  {Ashcroft}}\ and\ \bibinfo {author} {\bibfnamefont {N.~D.}\ \bibnamefont
  {Mermin}},\ }\href@noop {} {\emph {\bibinfo {title} {{Solid state
  physics}}}}\ (\bibinfo  {publisher} {Holt, Rinehart and Winston},\ \bibinfo
  {address} {New York},\ \bibinfo {year} {1976})\BibitemShut {NoStop}%
\bibitem [{\citenamefont {Lewenstein}\ \emph {et~al.}(1994)\citenamefont
  {Lewenstein}, \citenamefont {Balcou}, \citenamefont {Ivanov}, \citenamefont
  {L'Huillier},\ and\ \citenamefont {Corkum}}]{Lewenstein1994}%
  \BibitemOpen
  \bibfield  {author} {\bibinfo {author} {\bibfnamefont {M.}~\bibnamefont
  {Lewenstein}}, \bibinfo {author} {\bibfnamefont {P.}~\bibnamefont {Balcou}},
  \bibinfo {author} {\bibfnamefont {M.~Y.}\ \bibnamefont {Ivanov}}, \bibinfo
  {author} {\bibfnamefont {A.}~\bibnamefont {L'Huillier}}, \ and\ \bibinfo
  {author} {\bibfnamefont {P.~B.}\ \bibnamefont {Corkum}},\ }\href {\doibase
  10.1103/PhysRevA.49.2117} {\bibfield  {journal} {\bibinfo  {journal} {Phys.
  Rev. A}\ }\textbf {\bibinfo {volume} {49}},\ \bibinfo {pages} {2117}
  (\bibinfo {year} {1994})}\BibitemShut {NoStop}%
\bibitem [{\citenamefont {Lewenstein}\ \emph {et~al.}(1995)\citenamefont
  {Lewenstein}, \citenamefont {Sali{\`{e}}res},\ and\ \citenamefont
  {Lhuillier}}]{Lewenstein1995}%
  \BibitemOpen
  \bibfield  {author} {\bibinfo {author} {\bibfnamefont {M.}~\bibnamefont
  {Lewenstein}}, \bibinfo {author} {\bibfnamefont {P.}~\bibnamefont
  {Sali{\`{e}}res}}, \ and\ \bibinfo {author} {\bibfnamefont {A.}~\bibnamefont
  {Lhuillier}},\ }\href {\doibase 10.1103/PhysRevA.52.4747} {\bibfield
  {journal} {\bibinfo  {journal} {Phys. Rev. A}\ }\textbf {\bibinfo {volume}
  {52}},\ \bibinfo {pages} {4747} (\bibinfo {year} {1995})}\BibitemShut
  {NoStop}%
\bibitem [{\citenamefont {Milo{\v{s}}evi{\'{c}}}\ \emph
  {et~al.}(2006)\citenamefont {Milo{\v{s}}evi{\'{c}}}, \citenamefont {Paulus},
  \citenamefont {Bauer},\ and\ \citenamefont {Becker}}]{Milosevic2006}%
  \BibitemOpen
  \bibfield  {author} {\bibinfo {author} {\bibfnamefont {D.~B.}\ \bibnamefont
  {Milo{\v{s}}evi{\'{c}}}}, \bibinfo {author} {\bibfnamefont {G.~G.}\
  \bibnamefont {Paulus}}, \bibinfo {author} {\bibfnamefont {D.}~\bibnamefont
  {Bauer}}, \ and\ \bibinfo {author} {\bibfnamefont {W.}~\bibnamefont
  {Becker}},\ }\href {\doibase 10.1088/0953-4075/39/14/R01} {\bibfield
  {journal} {\bibinfo  {journal} {J. Phys. B At. Mol. Opt. Phys.}\ }\textbf
  {\bibinfo {volume} {39}},\ \bibinfo {pages} {R203} (\bibinfo {year}
  {2006})}\BibitemShut {NoStop}%
\end{thebibliography}

%

\end{document}